\documentclass[aps]{revtex4}
\usepackage{graphicx}

\def\bk{{\bf k}}

\newcommand{\lk}{\left( }
\newcommand{\rk}{\right)}
\newcommand{\ltk}{\left\{ }
\newcommand{\rtk}{ \right\} }
\newcommand{\ldk}{\left[ }
\newcommand{\rdk}{ \right] }

\def\betaT{{\tilde \beta}}
\def\muT{{\tilde \mu}}
\def\varepsilonT{{\tilde \varepsilon}}
\def\qB{{\bar q}}
\def\dB{{\bar d}}
\begin{document}

\title{Diquark Bose-Einstein condensation}

%\date{\today}

\author{K. Nawa\footnote{E-mail: nawa@ruby.scphys.kyoto-u.ac.jp}} 
\affiliation{Department of Physics, Kyoto University, 
                Kyoto 606-8502, Japan}

\author{E. Nakano\footnote{Current address: Department of Physics, National Taiwan University,
No.1, Sec.~4, Roosevelt Road, Taipei, Taiwan 106\\
E-mail: enakano@ntu.edu.tw}}
\affiliation{Yukawa Institute for Theoretical Physics, Kyoto University, 
              Kyoto 606-8502, Japan}
%\affiliation{Low temperature science research center, Kyoto University, 
%              Kyoto 606-8502, Japan}

\author{H. Yabu\footnote{E-mail: yabu@se.ritsumei.ac.jp}}
\affiliation{Department of Physics, Ritsumeikan University, Kusatsu, Shiga  
             525-8577, Japan} 
%\affiliation{Department of Physics, Tokyo Metropolitan University, 
%           1-1 Minami-Ohsawa, Hachioji, Tokyo 192-0397, Japan} 
%
\begin{abstract}
Bose-Einstein condensation (BEC) of composite diquarks 
in quark matter (the color superconductor phase) is discussed 
using the quasi-chemical equilibrium theory 
at a relatively low density region near the deconfinement phase transition,
where dynamical quark-pair fluctuations are assumed to be described 
as bosonic degrees of freedom (diquarks). 
A general formulation is given for the diquark formation and particle-antiparticle pair-creation processes 
in the relativistic framework, and some interesting properties are shown, 
which are characteristic for the relativistic many-body system.
Behaviors of transition temperature and phase diagram of the
quark-diquark matter are generally presented in model parameter space,
and their asymptotic behaviors are also discussed.
As an application to the color superconductivity, 
the transition temperatures and the quark and diquark density profiles
are calculated
in case with constituent/current quarks, 
where the diquark is in bound/resonant state. 
We obtained $T_C \sim 60-80\,{\rm MeV}$ for constituent quarks 
and $T_C \sim 130\,{\rm MeV}$ for current quarks at a moderate density ($\rho_b \sim 3 \rho_0$).
The method is also developed to include interdiquark interactions 
into the quasi-chemical equilibrium theory 
within a mean-field approximation, 
and it is found that a possible repulsive diquark-diquark interaction 
lowers the transition temperature by $\sim 50\,{\rm \%}$.

\end{abstract} 
\maketitle
%
%%%%%%%%%%%%%%%%%%%%%%%%%%%%%%%%%%%%%%%%%%%%%%%%%%%%%%%
\section{Introduction}
%%%%%%%%%%%%%%%%%%%%%%%%%%%%%%%%%%%%%%%%%%%%%%%%%%%%%%%
%

Several arguments based on the quantum chromodynamics (QCD) show
that the hadronic matter makes transition from the hadron phase 
to the quark-gluon one
at sufficiently high temperature and/or large baryon-number density. 
In particular, 
recent intensive studies on the quark matter 
suggest that, in low temperature, 
it becomes a condensed state of quark pairs 
called the color superconductor (CSC),  
where the color $SU(3)$ symmetry is spontaneously broken~\cite{BL1,ARW1,RSSV1}.  
In model calculations, 
the most attractive quark-quark interaction 
induced by gluon exchange
appears in the total spin singlet ($J=0$) 
and the color anti-triplet (${\bar {\bf 3}}$) channel,
so that the CSC phase 
is considered to be a state 
where the quark pairs with these quantum numbers 
are condensed coherently 
in a similar mechanism with the BCS theory~\cite{BCS1}. 

Generally, 
there exist two limiting cases in the fermion-pair condensation: 
weak- and strong-coupling regions 
in the strength of the two-fermion correlation.  
In the weak-coupling region, 
which appears in the high-density quark matter 
because of the asymptotic-freedom of QCD~\cite{CP1}, 
the coherence length $\xi$ becomes larger 
than the interquark distance $d$, 
and the quark matter becomes
the weakly-interacting fermion system 
where the mean-field picture are applicable.
The strong-coupling region appears 
in the moderate- or low-density quark matter 
(especially in the neighbor of the deconfinement transition point).
Large pair-fluctuations develop~\cite{Kitazawa} 
and the coherence length is suggested to be very small;
$\xi/d \sim O(1-10)$, even around the moderate-density region [$\sim
O(10)\rho_0$] from the analysis of the improved ladder approximation~\cite{AHI1}.   
This implies that, for low-density region [$\sim O(1)\rho_0$], $\xi/d
\ll 1$ and the strong quark-pair correlations 
are described by 
bosonic degrees of freedom, 
which we call ``diquark'' in this paper.
Thus, for the description of this diquark region of the quark matter, 
the incorporation of two-particle correlations beyond mean-field levels 
is essentially important.

Several methods have been proposed 
to incorporate the correlation effects. 
A typical one is that of 
Nozi\`{e}res and Schmitt-Rink, 
where thermodynamic potentials are calculated 
in the normal phase using the in-medium ladder approximation, 
and the fermion number density is shown to include 
a contribution from the bosonic part 
through the two-body in-medium scatterings~\cite{NSR1, Rand1}.
The transition temperature is obtained 
from the zero-frequency pole of the two-particle correlation function 
(Thouless criterion for second-order phase transitions). 

The quasi-chemical equilibrium theory 
is an another approach   
developed by Schafroth, Butler, and Blatt~\cite{SBB1}
for the fermion-pair condensates.
Using a cluster expansion technique
for the partition function, 
the fermion-pair condensates are shown to be 
in chemical equilibrium states 
between the isolated fermions and the bosonic difermions 
identified with the in-medium resonant/bound states of two fermions. 
This method was firstly invented to describe the experimental phenomena
of electric superconductivity of metals in 1940', before the birth of BCS
theory. In modern sense, this quasi-chemical equilibrium picture
corresponds to the strong coupling superconductor ($\xi/d \leq 1$), and
the transition temperature, 
which is obtained as the BEC-$T_C$ of difermions 
under the equilibrium, 
is shown to be equivalent with that in Nozi\`{e}res-Schmitt-Rink theory 
in the strong-coupling limit ($\xi/d \ll 1$).

In the strong-coupling region of the quark matter, 
we can expect that 
the similar quark-pair fluctuations develop 
and the CSC is described by the diquark condensate 
in equilibrium with isolated quarks.

The diquark correlation effects have been proposed 
by Jaffe and Low~\cite{Jaff1, JaffLow1} 
in relation with the phenomenology of 
the low-energy hadron-hadron collisions and 
the multi-hadron spectroscopy. 
Also, roles of composite diquarks in the quark-gluon plasma 
have been studied by Ekelin, et al.~\cite{Eke1, Eke2}, 
especially for the  hadronization process 
in the high-energy $e^+e^-$ collision; 
they considered the space-time developments 
of the quark-gluon plasma 
as an adiabatic process, 
and discussed the possibility of the diquark BEC 
rather at high-$T$ stages in that process.
Also the diquark correlations have been 
discussed to study the stability of diquark matter
but without regard to diquark BEC~\cite{DS1, Horv1, Sat1, KKM1, KT1}. 

In this paper, 
we make a general formulation of the quasi-chemical equilibrium theory 
for the relativistic many-fermion system, 
and apply it to the strong-coupling quark matter with composite diquarks. 
Our special interest is on diquark BEC 
that should be the CSC in the strong-coupling region, 
and we clarify the CSC states as diquark BEC 
at finite temperature and density 
within the quasi-equilibrium theory 
in less model-dependent way. 
In Sec.~II, 
we formulate the quasi-chemical equilibrium theory 
for the difermion formation 
and the particle-antiparticle pair-creation processes 
in the relativistic many-body system.
We also show some interesting properties for the BEC of difermions 
which are characteristic in the relativistic system. 
In Sec.~III, applying the method to the quark-diquark matter, 
we show $T_C$ of the diquark BEC, a phase diagram on quark/diquark masses, and
asymptotic behaviors of them.
We also discuss $T_C$ of the color superconductor
using phenomenological values of parameters.
In Sec.~IV, 
effects of the interdiquark interactions 
are discussed in a mean-field approximation.
Sec.~V is devoted to summary and outlook. 
Throughout this paper, we use the natural units: 
$c=\hbar=k_B=1$.

%%%%%%%%%%%%%%%%%%%%%%%%%%%%%%%%%%%%%%%%%%%%%%%%%%%%%%%%%%%%%%%%%%%%%
\section{Relativistic quasi-chemical equilibrium theory \label{Quasi}}
%%%%%%%%%%%%%%%%%%%%%%%%%%%%%%%%%%%%%%%%%%%%%%%%%%%%%%%%%%%%%%%%%%%%% 

%%%%%%%%%%%%%%%%%%%%%%%%%%%%%%%%%%%%%%%%%%%%%%%%%%
\subsection{Molecular formation/annihilation process}
%%%%%%%%%%%%%%%%%%%%%%%%%%%%%%%%%%%%%%%%%%%%%%%%%%

To develop the quasi-chemical equilibrium theory, 
we consider molecular formation/annihilation process
in a fermionic matter:
\begin{eqnarray}
     F_1 +F_2 \longleftrightarrow (F_1 F_2)=B,  
\label{equilib1}  
\end{eqnarray}
where $F_{1,2}$ are fermions with masses $m_F \equiv m_{F_1} =m_{F_2}$, 
and $B$ is a composite boson of $F_1$ and $F_2$ 
with mass $m_B$; 
the composite boson $B$ should be a bound state ($m_B < 2m_F$)  
or a resonance state ($m_B > 2m_F$). 
We call it ``molecule'' in both cases.

Here we take the quasi-particle picture 
for the fermionic matter where
the system consists of quasi-fermions ($F_1$ and $F_2$)
and quasi-bosons ($B$).
The two-body interactions between fermions  
bring about the two-body correlations  
in the matter; 
their major effects are for the binding/resonance energy 
of the composite boson $B$ 
and the residual interactions among quasi-particles.
This quasi-particle interactions are generally regarded to be weak, 
and we neglect it in the first part of this paper. 
In Sec.~IV, we introduce the quasi-particle interactions
and discuss their effects within a mean-field approximation.

The equilibrium condition for the process (\ref{equilib1}) 
is given by 
\begin{equation}
     \mu_{F_1} +\mu_{F_2} =\mu_B,
\label{equilib2}  
\end{equation}
where $\mu_{F_1,F_2,B}$ are the chemical potentials 
of the particles $F_1$, $F_2$ and $B$, 
which should be functions of the temperature $T$ 
and the particle-number densities $n_{F_1,F_2,B}$.  
For free uniform gases, 
they are given by the Bose/Fermi statistics:
\begin{eqnarray}
     n_B &=& \frac{1}{(2\pi)^3} 
             \int\frac{d^3\bk}{e^{\beta (\varepsilon_B -\mu_B)} -1},  
\label{bose1} \\ 
     n_{F_\alpha} &=& \frac{1}{(2\pi)^3} 
                      \int\frac{d^3\bk}{e^{\beta (\varepsilon_{F_\alpha} 
                                                  -\mu_{F_\alpha})} +1},  \quad
     (\alpha=1,2)
\label{fermi1}
\end{eqnarray}
where $\beta=1/T$. 
Using the relativistic
energy-momentum dispersion relations: 
$\varepsilon_a =\sqrt{ m_a^2 +\bk^2}$ $(a=F_1,F_2,B)$, 
eqs. (\ref{bose1},\ref{fermi1}) become
\begin{eqnarray}
     n_B &=& \frac{1}{2 \pi^2} 
             \int_{m_B}^\infty
             \frac{\varepsilon \sqrt{\varepsilon^2 -m_B^2}}{ 
                    e^{\beta(\varepsilon -\mu_B)} -1}
                   d\varepsilon
         \equiv m_B^3 B_R(-\beta \mu_B,m_B \beta),
\label{bose2} \\ 
     n_{F_\alpha} &=& \frac{1}{2 \pi^2} 
                      \int_{m_{F_\alpha}}^\infty
                      \frac{\varepsilon 
                            \sqrt{\varepsilon^2 -m_{F_\alpha}^2}}{ 
                    e^{\beta(\varepsilon -\mu_{F_\alpha})} +1}
                    d\varepsilon
                   \equiv m_{F_\alpha}^3 F_R(-\beta \mu_{F_\alpha},m_{F_\alpha} \beta),  
     \qquad
     (\alpha=1,2)
\label{fermi2}
\end{eqnarray}
In the above equations, 
the relativistic Bose/Fermi functions $B_R$ and $F_R$ are defined by
\begin{eqnarray}
     B_R(\nu,\betaT) &=& \frac{1}{2 \pi^2} 
                         \int_1^\infty
                         \frac{\varepsilonT \sqrt{\varepsilonT^2 -1}}{ 
                               e^{\betaT \varepsilonT +\nu} -1}
                          d\varepsilonT
                      = \frac{1}{2 \pi^2} 
                        \int_0^\infty
                        \frac{\cosh\theta (\sinh\theta)^2}{ 
                              e^{\betaT \cosh\theta +\nu} -1}
                          d\theta,
\label{bose3} \\ 
     F_R(\nu,\betaT) &=& \frac{1}{2 \pi^2} 
                         \int_1^\infty
                         \frac{\varepsilonT \sqrt{\varepsilonT^2 -1}}{ 
                               e^{\betaT \varepsilonT +\nu} +1}
                          d\varepsilonT
                      = \frac{1}{2 \pi^2} 
                        \int_0^\infty
                        \frac{\cosh\theta (\sinh\theta)^2}{ 
                              e^{\betaT \cosh\theta +\nu} +1}
                          d\theta,
\label{fermi3}
\end{eqnarray}
where we use scaled variables  
$\betaT =m \beta$ and $\varepsilonT =\varepsilon/m$, 
and the J\"uttner variable $\theta$ that is defined 
by $\varepsilonT =\cosh\theta$~\cite{Juttner,Chandra}. 

The boson chemical potential $\mu_B$ satisfies 
$\mu_B \leq m_B$ from the positive norm condition of $n_B$, 
and the boson density (\ref{bose2}) has a singularity 
at this upper bound, 
where a phase transition occurs to the Bose-Einstein condensate. 
In the BEC region ($\mu_B=m_B$), 
the boson density consists of the condensed and thermal parts, 
$n_B =n_B^{(0)} +n_B^{(th)}$, 
which are given by 
$n_{B}^{(th)} =m_B^3 B_R(-\beta m_B,m_B \beta )$ 
and $n_B^{(0)} =n_B -n_B^{(th)}$~\cite{Lond1}.

In the process (\ref{equilib1}), 
the particle-number conservations for $F_{1,2}$ give constraints for the densities:
\begin{eqnarray}
     n_{F_1} +n_B =n_{F_1,t},  \qquad
     n_{F_2} +n_B =n_{F_2,t},  
\label{densitycons}  
\end{eqnarray}
where $n_{F_{1,2},t}$ are the total number densities 
of $F_{1,2}$, which consist of isolated fermions and those included in
the composite boson $B$.
%$n_{F_{1,2},t}$ should be conserved through the process  (\ref{equilib1}).

Solving eqs. (\ref{equilib2}) and (\ref{densitycons}) 
with (\ref{bose2},\ref{fermi2}), 
we can obtain the densities $n_{B,F_1,F_2}$ in equilibrium at temperature $T$.

%%%%%%%%%%%%%%%%%%%%%%%%%%%%%%%%%%%%%%%%%%%%%%%%%%
\subsection{Particle-antiparticle pair-creation/annihilation process}
%%%%%%%%%%%%%%%%%%%%%%%%%%%%%%%%%%%%%%%%%%%%%%%%%%

The energy scale in the quark-diquark system is so large 
that the antiparticle degrees of freedom are generally important.
Here we develop the quasi-chemical equilibrium theory 
for the particle-antiparticle system. 

The fermion-antifermion (quark-antiquark) or 
boson-antiboson (diquark-antidiquark) pairs can be created 
through the radiation process mediated by the gauge boson (gluon); 
\begin{eqnarray}
     G \longleftrightarrow P +{\bar P},  \qquad
     (P =F_{1},F_{2},B)
\label{equilib3}  
\end{eqnarray}
Thus 
the equilibrium condition for the above process 
is given by
$\mu_P +\mu_{{\bar P}} =\mu_G$, 
where $\mu_{P,{\bar P},G}$ are the chemical potentials 
of the particle $P$, the antiparticle ${\bar P}$ and the gluon $G$. 
We put $\mu_G=0$ because the gluon is a massless particle, 
so that the chemical potentials of $P$ and ${\bar P}$ 
have a relation: $\mu_{{\bar P}} =-\mu_P$. 

In the process (\ref{equilib3}), 
the difference between the particle and antiparticle densities 
should be conserved:
\begin{eqnarray}
     n_{b,P} =n_P -n_{{\bar P}},  \qquad
     (P=F_1,F_2,B)
\label{bdensitycons}  
\end{eqnarray}
where $n_{b,P}$ corresponds to the baryon-number density 
of quark or diquark degrees of freedom.
As given in (\ref{bose2},\ref{fermi2}),
the boson/fermion number-densities of the particle and antiparticle 
at the temperature $T$ are given by
\begin{eqnarray}
     n_B          &=m_B^3 B_R(-\beta \mu_B,m_B \beta), \qquad
     n_{{\bar B}} &=m_B^3 B_R(\beta \mu_B,m_B \beta),  
\label{bose4} \\ 
     n_{F_\alpha} &=m_{F_\alpha}^3 F_R(-\beta \mu_{F_\alpha},m_{F_\alpha} \beta), \qquad
     n_{{\bar F_\alpha}} &=m_{F_\alpha}^3 F_R(\beta
     \mu_{F_\alpha},m_{F_\alpha} \beta),
     \qquad
     (\alpha=1,2) 
\label{fermi4} 
\end{eqnarray}
where $B_R$ and $F_R$ have been defined in (\ref{bose3},\ref{fermi3}).

Eq. (\ref{bose2}) shows that 
the boson chemical potentials are found to have the upper bound; 
$\mu_{B,{\bar B}} \leq m_B$. 
From the relation $\mu_B =-\mu_{{\bar B}}$, 
they are also found to have the lower bound:
\begin{equation}
     -m_B \leq \mu_{B,{\bar B}} \leq m_B,
\label{conditionmud1}
\end{equation}
where $\mu_B \geq 0$ ($\mu_B \leq 0$) corresponds to 
$n_{b,B} =n_B-n_{{\bar B}} \geq 0$ ($n_{b,B}\leq 0$).
Furthermore, 
the upper and lower bounds of $\mu_B$ in (\ref{conditionmud1}) 
correspond to the boson BEC ($\mu_B =m_B$) and antiboson BEC 
($\mu_{\bar B} =m_B$) respectively,
which means that bosons and antibosons cannot be condensed to the BEC state
at the same time in equilibrium states.

The transition temperature $T_C$ of the boson BEC (the antiboson BEC) 
is determined by (\ref{bdensitycons}) with $\mu_B =m_B$ ($\mu_B =-m_B$):
\begin{equation}
     n_{b,B} =[ n_B -n_{{\bar B}} ]_{\mu_B =\pm m_B,T=T_C}. 
\label{ttba}
\end{equation}

%%%%%%%%%%%%%%%%%%%%%%%%%%%%%%%%%%%%%%%%%%%%%%%%%%%%%%%%%%%%%%%%%%%%%
\subsection{Molecular formation/annihilation processes of many kinds of particles}
%%%%%%%%%%%%%%%%%%%%%%%%%%%%%%%%%%%%%%%%%%%%%%%%%%%%%%%%%%%%%%%%%%%%%

Here we consider a general system of many kinds of fermions $F_\alpha$ 
and molecular bosons $B_\beta$ which are bound/resonant states of two fermions.  
The fermions are assumed to be transmuted 
through some processes; $F_\alpha \leftrightarrow F_{\alpha'}$.
As is same with the previous simple case, the total number density of
fermions $n_{F,t}$ is to be conserved in molecular formation/annihilation
process of $B_\beta$, and is given by summing all species of fermions
$n_{F_\alpha,t}$, which also amounts to summation of isolated fermions $n_{F_\alpha}$
and those included in composite bosons $n_{B_\beta}$:
\begin{equation}
     n_{F,t} \equiv \sum_\alpha n_{F_\alpha,t}
             =  \sum_\alpha n_{F_\alpha}
             +2 \sum_\beta  n_{B_\beta}.
\label{subtract1}
\end{equation}
To obtain the equilibrium condition for the above condition,  
we take the Helmholtz free-energy density of the system:
\begin{eqnarray}
     f &\equiv& F/V 
        = f_{T, V}
         +\sum_\alpha \mu_{F_\alpha} n_{F_\alpha}
         +\sum_\beta  \mu_{B_\beta}  n_{B_\beta} 
\nonumber\\
       &&+K \lk n_{F,t} -\sum_\alpha n_{F_\alpha} -2 \sum_\beta  n_{B_\beta}\rk,
\label{mixed_energy1}
\end{eqnarray}
where $f_{T,V}$ is the density-independent part, 
and the last term has been added 
to expose the constraint (\ref{subtract1}) 
with the Lagrange multiplier $K$.
Differentiating eq. (\ref{mixed_energy1}) 
with respect to $n_{F_\alpha}$ and $n_{B_\beta}$, 
one obtains for ${}^\forall\alpha, {}^\forall\beta$ 
\begin{equation}
     2 \mu_{F_\alpha} = 2 K =\mu_{B_\beta}.
\label{derivative1}  
\end{equation}
This means that the chemical potentials of fermions and bosons
take the same values $\mu_F$ and $\mu_B$ in an equilibrium state, 
which obey the condition (\ref{equilib2}); $2\mu_F=\mu_B$.

The range of the chemical potential is 
$\mu_B =\mu_{B_\beta} \leq m_{B_\beta}$ 
for any $B_\beta$ and, 
without loss of generality, 
the boson $B_1$ is assumed to be the one 
with the smallest mass (most-stable composite); 
$m_{B_1} < m_{B_{\beta\neq 1}}$. 
Then, the range of $\mu_B$ becomes
\begin{equation}
     \mu_B (=2 \mu_F) \leq m_{B_1}.
\label{rangeB}  
\end{equation}
It shows that, 
at the upper bound of the chemical potential 
($\mu_{B_1} =\mu_B =m_{B_1}$), 
the BEC occurs for the boson $B_1$, 
but there exists no BEC singularity for the other bosons 
because $\mu_{B_{\beta \neq 1}} =\mu_B =m_{B_1} < m_{B_\beta}$. 
In summary, 
the BEC occurs only for the boson with the smallest mass 
below the transition temperature $T_C$,
and the other bosons with larger masses (less-stable composites) cannot be condensed 
to the BEC states 
at any temperature and density (one-BEC theorem).

The critical temperature $T_C$ for the $B_1$'s BEC  
can be determined from the condition at $T=T_C$:
\begin{equation}
     \ldk \sum_\alpha n_{F_\alpha} 
         +2 \sum_\beta n_{B_\beta}
     \rdk_{2\mu_F=\mu_B \rightarrow m_{B_1}} 
     = n_{Ft}. 
\label{upper3}
\end{equation}

Below $T_C$, 
the density of $B_1$ consists of the thermal and condensed parts:
\begin{eqnarray}
     n_{B_1}^{(th)} &=& m_{B_1}^3 B_R(-\beta m_{B_1},m_{B_1} \beta),  
\label{b1nc}\\
     n_{B_1}^{(0)}  &=& \frac{1}{2} 
                   \left[ n_{F,t} -\sum_\alpha n_{F_\alpha} 
                                  -2 \sum_{\beta \neq 1} n_{B_\beta} 
                   \right] -n_{B_1}^{(th)},
\label{b1c}
\end{eqnarray}
where $n_{F,t} =\sum_\alpha n_{F_\alpha,t}$ is the total fermion number, 
and $n_{B_\beta,F_\alpha}$ are given by (\ref{bose4},\ref{fermi4}) again.

In the case of the system with antiparticle degrees of freedom,
antibosons cannot be condensed to BEC state if the baryon-number density
is positive.
This can be also understood by one-BEC theorem.
The ground state contributions of bosons ($B_1$) and antibosons (${\bar
B}_1$) are explicitly given by
\begin{eqnarray}
     n_{B_1}^{(0)} &=& \frac{1}{V} 
             \frac{1}{e^{\beta (m_{B_1} -\mu_B)} -1},\\ 
     n_{\bar{B}_1}^{(0)} &=& \frac{1}{V} 
             \frac{1}{e^{\beta (m_{B_1} +\mu_B)} -1}%
                          =  \frac{1}{V} 
             \frac{1}{e^{\beta ((m_{B_1}+2\mu_B) -\mu_B)} -1},
 \end{eqnarray}
 which means that the antiboson is effectively heavier than boson by
 $2\mu_B (>0)$ and BEC occurs only for bosons $B_1$ 
by using one-BEC theorem 
(In case that the baryon-number density is negative, $\mu_B<0$, 
only $\bar{B}_1$ can then be condensed to BEC state
at the lower bound $\mu_B=-m_{B_1}$).  
%%%%%%%%%%%%%%%%%%%%%%%%%%%%%%%%%%%%%%%%%%%%%%%%%%%%%%%%%%%%%%%%%%%%%
\section{Diquark Bose-Einstein condensation in QCD}
%%%%%%%%%%%%%%%%%%%%%%%%%%%%%%%%%%%%%%%%%%%%%%%%%%%%%%%%%%%%%%%%%%%%%

In this section,
we apply the relativistic quasi-chemical equilibrium theory developed 
in the previous section 
to the color superconductivity of quark matter in BCS-BEC
crossover regime \cite{Egor, N_Abu}, 
where quark-pair fluctuations develop dynamically to create diquarks (composite bosons). 
A possibility of quark-quark correlations has also been suggested 
in low-energy hadron physics \cite{Jaff1, JaffLow1}, 
and its extrapolation to the deconfinement phase 
can be considered to become diquark degrees of freedom in dense region 
\cite{DS1, Horv1, KKM1}. 

%%%%%%%%%%%%%%%%%%%%%%%%%%%%%%%%%%%%%%%%%%%%%%%%%%%%%%%%%%%%%%%%%%
\subsection{Applicability of quasi-chemical equilibrium theory for quark
  matter}
%%%%%%%%%%%%%%%%%%%%%%%%%%%%%%%%%%%%%%%%%%%%%%%%%%%%%%%%%%%%%%%%%%

Before going to actual calculations 
we should evaluate the applicability of 
quasi-chemical equilibrium theory for interacting fermionic systems, 
which was originally discussed by Schafroth, et al. with 
respect to electron gas in superconducting metals \cite{SBB1}: 
 
\begin{list}{}{}
\item[1)] particles of compositions 
(fermions $F$ and composite molecules $B$) 
should be quasi-particles, 
\item[2)] correlations between $FB$ and $BB$ are unimportant,
\item[3)] higher-order correlations are not stronger than the pair-correlations 
considered, 
\item[4)] any localizations to liquid droplets with large volume exclusion 
(gas-liquid transition) do not occur before the formation of 
quasi-molecules,
\item[5)] composites sufficiently obey the Bose statistics 
as expected with small size of composites 
relative to the interparticle distance. 
\end{list}

Now we discuss the condition 1) for quark matter. 
The lifetime of a particle relates to various decay processes in medium, 
and to recognize quark as quasi-particle 
its lifetime should be at least longer than 
inverse of the QCD energy scale, that is, $O(1)$ fm. 
The lifetime of quarks tends to be shorter near the Fermi surface,
which can be roughly estimated from the imaginary part 
of quark self-energy, e.g., \cite{KitaL}; $\tau \sim 30^{-1}$ MeV$^{-1}$.
In our model these quarks 
with short lifetime near Fermi surface due to strong pair fluctuations 
are described as bosonic molecules, 
and residual quarks in the deeper interior of the Fermi sphere 
have explicitly longer lifetime than the QCD scale. 
Thus, quarks (antiquarks) being outside of pairs due to medium effects 
can be identified as quasi-particles. 
As for the diquark composites, 
its lifetime as quasi-particle is expected 
to be longer than that of quarks in crossover regime, 
as Shuryak and Zahed suggested 
the existence of 
various binary bound states 
in the whole regions just above the deconfinement temperature 
because QCD coupling strength is found not to be so small as one has believed so far 
\cite{shur}. 
Thus, diquarks should also be identified in medium. 

Concerning other conditions 2) and 3), 
higher-order correlations rather than pair ones are not treated in
this Sec.~III. 
This treatment corresponds to the Gaussian-type approximations 
like Nozi\`{e}res and Schmitt-Rink approach \cite{NSR1,Rand1,N_Abu}.
The effects of higher-order correlations, 
especially interdiquark interactions, will be discussed in
Sec.~IV with sincere generalization of quasi-chemical equilibrium
theory. Condition 4) is also discussed in the end of Sec.~IV.

With regard to condition 5), the density dependence of coherence length
$\xi$, characterising typical size of a Cooper pair, relative to
interquark distance $d$ is estimated by using the improved ladder
approximation~\cite{AHI1},
which suggests that $\xi/d \leq 1$ for $\rho_b<10\rho_0$.
This means that $\xi/d \ll 1$ for lower density region $\rho_b \sim 3\rho_0$
concerned,
and our quasi-chemical equilibrium approach with bosonic modes is
expected to work well for these strong couplig regimes of quark matter.

The critical coupling strength for the formation of diquark bound states
is also calculated by using effective four-Fermi interaction model with
Gaussian approximation~\cite{N_Abu},
and critical value $G/(\pi^2/\Lambda^2)\simeq 0.92$ ($\Lambda $ is
ultraviolet cutoff scale) and its relavance for real quark matter is
suggested therein. 
Because of smooth crossover, 
this critical value of coupling strength
corresponding to the unitary limit in cold atomic systems
brings no sudden changes 
for physical quantities. 
The essential quantity
for the behavior of BEC-BCS crossover
is always 
the ratio of the coherence length to the interparticle distance.  

Although above quantitative argumets come from some specific model calculations, 
recent analyses on QCD experiments, 
e.g., relativistic heavy ion collision~\cite{HIC} and lattice QCD~\cite{latticeQCD}, 
also suggest that even in deconfinement phase 
QCD holds strong coupling 
at almost whole region of the phase diagram relevant to possible observations. 
We expect that our present approach works well in such strong coupling regimes 
in more general way, 
especially around the region close to the phase transition line 
where correlations between quarks should develop largely and 
the strong coupling picture becomes essential for the pairing problem. 
In this context three-quark correlations in addition to pairing 
will be important issue around this region, 
which can also be treated in our approach. 

Here we comment on the BEC-BCS crossover in beta-equilibriated  
charge-neutral quark matter,
where mismatches of Fermi surfaces exist in different flavors.
Even in the weak coupling regime,
various phases have been proposed in that matter in relation with the  
chromomagnetic instability problem:
the gapless superconductor~\cite{gapless}, 
the crystalline color superconductor~\cite{crystalline}, 
the gluon condensed state~\cite{Hashi}, and so on.
The strong coupling limit has also been discussed of the two-flavor  
quark matter with Fermi-surface mismatch~\cite{KitaRS}.
These considerations suggest that the Cooper-pair formation should still be  
favored in the case
if the energy gain by the pairing exceeds the energy differences of  
the mismatched Fermi surfaces.
Thus we can expect the crossover into the diquark BEC in the quark  
matter with finite mismatches
for sufficiently large coupling strength and,
at the temperature around $T_C$,
the mismatch effects are expected to be small 
with large pairing energy.
In low temperature region, however,
the Fermi-surface mismatches makes
the unpaired fermions remain with some flavor,
and they should form new Fermi surfaces;
the calculation of the profile changes of them around the crossover  
region should be a very interesting problem
in future.

In the present paper, we take the two-flavor quark matter with  
same Fermi surfaces
to discuss typical features of diquark BEC;
however, our approach can be extended to the case of the finite Fermi- 
surface mismatches
and  their effects in the diquark BEC and crossover will be discussed in  
another paper.

%%%%%%%%%%%%%%%%%%%%%%%%%%%%%%%%%%%%%%%%%%%%%%%%%%%%%%%%%%%%%%%%%%
\subsection{Baryon-number conservation and chemical equilibrium condition}
%%%%%%%%%%%%%%%%%%%%%%%%%%%%%%%%%%%%%%%%%%%%%%%%%%%%%%%%%%%%%%%%%%

In what follows, we take u,d-quark matter around the moderate density
region in QCD, 
and consider the chemical equilibrium between quarks and 
composite diquarks in the 2SC-type pairing state
(color antisymmetric ${\bar {\bf 3}}$, 
flavor antisymmetric, and total-spin $J=0$)~\cite{BL1, ARW1, RSSV1}. 
The other types of diquarks can also be created in the quark matter
using less attractive channels, 
e.g., diquarks in color-${\bar {\bf 3}}$, 
flavor symmetric and the $J=1$ pairing state~\cite{NMT1, SWR1}. 
But, according to the one-BEC theorem derived in the previous section, 
the diquark BEC can occur only for the smallest mass (most stable) diquark, 
i.e. in the 2SC-diquark state of u and d quarks, 
so that we consider only the 2SC-diquarks.
Generally the diquarks which do not participate in the BEC 
would have the effect of lowering the critical temperature, 
through the total-number conservation,
in comparison with the case of one kind of diquark.
We also assume that the quark and diquark masses do not depend 
on the temperature and density, 
but the present formalism is also applicable to the case 
with effective masses dependent on these parameters.

In the quark-diquark matter, 
the baryon-number density $\rho_b (\equiv n_b/3)$ 
corresponds to the conserved total fermion-number density $n_{F,t}$ 
in (\ref{bdensitycons}) or (\ref{subtract1}), 
where $n_b$ is the regularized baryon-number density in quark degrees. 
In the chemical equilibrium between quarks and 2SC-diquarks,  
it is given by
\begin{equation}
     n_b =3\cdot 2\cdot 2\cdot \left( n_q -n_{\qB} \right)
         +2 \cdot 3\cdot 1\cdot 1\cdot \left( n_d -n_{\dB} \right), 
\label{QCD_cons1}
\end{equation}
where $n_{q,\qB}$ are the number densities of quark and antiquark, 
and $n_{d,\dB}$ are those of diquark and antidiquark.
The numerical factors in (\ref{QCD_cons1}) come from 
the degeneracy of the internal degrees of freedom:
the color-triplet (3), the flavor-doublet (2), the spin-doublet (2) 
for the quark-density term, 
and the compositeness (2), the color-anti-triplet (3), the flavor-singlet (1), 
the spin-singlet (1) for the diquark-density term. 

Eqs.~(\ref{bose4},\ref{fermi4}) gives the densities $n_{q,\qB}$ 
and $n_{d,\dB}$ in equilibrium at temperature $T$:
\begin{eqnarray}
     n_{q}   &=m_q^3 F_R(-\beta \mu_q,m_q \beta),  \qquad
     n_{\qB} &=m_q^3 F_R(\beta \mu_q,m_q \beta),  
\label{QCD_fermidense} \\ 
     n_d     &=m_d^3 B_R(-\beta \mu_d,m_d \beta),  \qquad
     n_{\dB} &=m_d^3 B_R(\beta \mu_d,m_d \beta),  
\label{QCD_bosedense} 
\end{eqnarray}
where $m_{q,d}$ are a quark/antiquark and a diquark/antidiquark masses, 
and $\mu_{{\bar P}} =-\mu_P$  ($P=q,d$) has been used for the chemical potentials 
of antiquark and antidiquark.
When the diquark (antidiquark) BEC occurs for $T \leq T_C$, 
the diquark (antidiquark) density consists of the condensed and thermal parts: 
$n_d =n_d^{(0)} +n_d^{(th)}$ ($n_{\dB} =n_{\dB}^{(0)} +n_{\dB}^{(th)}$), which are obtained 
in the same way as eqs. (\ref{b1nc},\ref{b1c}). 
Especially these condensates $n_d^{(0)}$ and $n_{\dB}^{(0)}$ are the
superfluid density of CSC.
If the baryon-number density is positive, diquark BEC can only be
expected as mentioned in the previous section.

The chemical potentials $\mu_q$ and $\mu_d$ should obey 
the quark-diquark equilibrium condition:
\begin{equation}
     2 \mu_q =\mu_d, 
\label{QCD_chemical_eq4}
\end{equation}
which corresponds to (\ref{equilib2}). 
The set of eqs.~(\ref{QCD_cons1}-\ref{QCD_chemical_eq4}) 
determines a composition of the baryon-number density 
in the quark-diquark quasi-chemical equilibrium system. 

%%%%%%%%%%%%%%%%%%%%%%%%%%%%%%%%%%%%%%%%%%%%%%%%%%%%%%%%%%%%%%%%%%%%%%
\subsection{Transition temperature of quark-diquark system}
%%%%%%%%%%%%%%%%%%%%%%%%%%%%%%%%%%%%%%%%%%%%%%%%%%%%%%%%%%%%%%%%%%%%%%

As shown in the previous section, 
diquark BEC occurs 
when the diquark chemical potential reaches 
the upper limit $\mu_d = m_d$.
In the quark-diquark system
considered here,
the conditions for $T_C$ are 
\begin{eqnarray}
     2 \mu_q &=& \mu_d =m_d, 
\label{QCD_critical_mixed0}\\
     n_b &=& \ldk {3\cdot 2\cdot 2 \cdot} (n_q -n_{{\bar q}}) 
            +2 \cdot {3\cdot 1\cdot 1 \cdot} 
                         (n_d^{(th)}-n_{\dB}^{(th)}) 
                     \rdk_{T=T_C},  
\label{QCD_critical_mixed1}
\end{eqnarray}
where we have used eqs.~(\ref{QCD_cons1},\ref{QCD_chemical_eq4}).
For any fixed values of the quark and diquark masses $m_{q,d}$, 
and for the baryon-number density $n_b$ in (\ref{QCD_critical_mixed1}), 
$T_C$ is obtained by solving these equations.

\begin{figure}[ht]
  \begin{center}
    \begin{tabular}{cc}
       \resizebox{70mm}{!}{\includegraphics{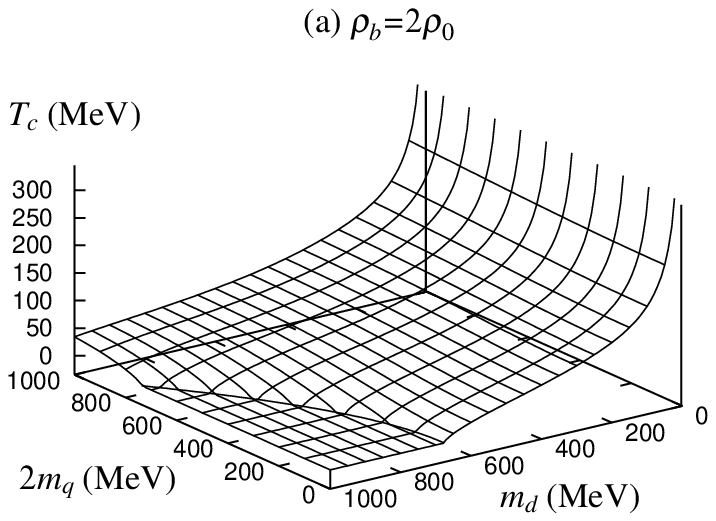}} &
       \resizebox{70mm}{!}{\includegraphics{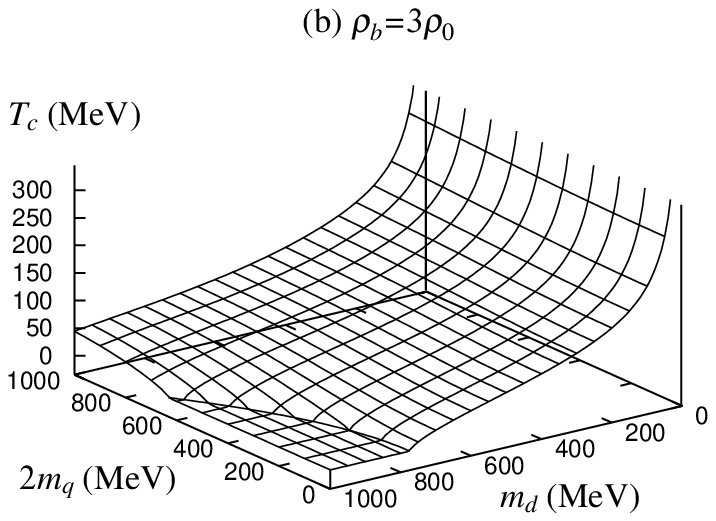}}
    \end{tabular}\par
\vspace{-5mm}
       \resizebox{70mm}{!}{\includegraphics{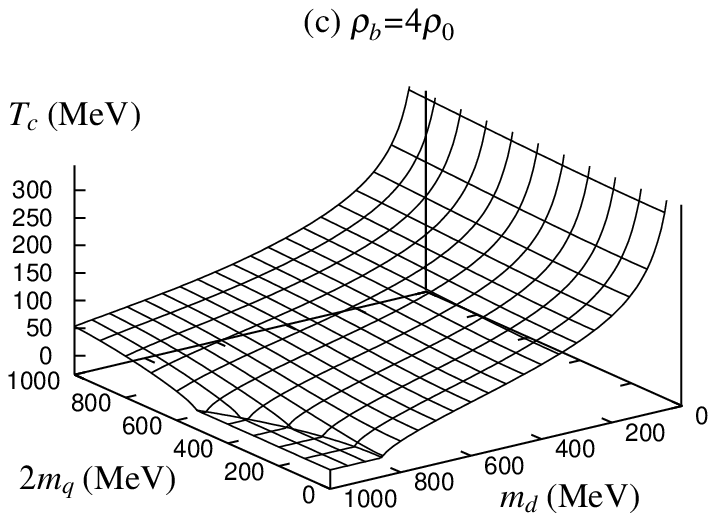}}\par
  \end{center}
\caption{Transition temperatures $T_C$ of diquark BEC 
         on $(2 m_q,m_d)$-plane of quark/diquark masses 
         for the baryon-number densities $\rho_b =2 \rho_0, 3 \rho_0, 4 \rho_0$ 
         ($\rho_0$ is the normal nuclear density).}
\label{fig1}
\end{figure}

Fig.~\ref{fig1} shows
dependence of $T_C$ on $2 m_q$ and $m_d$
for the baryon-number densities  
$\rho_b=2 \rho_0, 3 \rho_0, 4 \rho_0$ 
where $\rho_0 \sim 1.3 \times 10^6\,{\rm MeV^3}$ is the normal nuclear density.
These figures show that $T_C$ increases 
with the baryon-number density;
especially it can be read 
from the shrinkage of no-BEC region
(area in the front side in the $(2 m_q$,$m_d)$-plane).
This effect can be understood 
that $T_C$ goes up with increase of the diquark density 
together with the baryon density, 
as seen in the behavior of $T_C$ in the ideal bose gases 
(see Appendix~\ref{AppB}). 

%---------------
%--------------

These diagrams also show 
that the large diquark mass 
has an effect of lowering $T_C$. 
It is consistent with the results 
for one-component boson-antiboson gas (see Appendix~B);
$T_C \propto m_d^{-1}$ in nonrelativistic limit (\ref{ApA8}) and
$\propto m_d^{-1/2}$ in ultrarelativistic limit (\ref{ApA16}).
The increase of $T_C$ for large quark-masses is attributed to 
the large diquark density  
brought by the increase in the binding energy 
($m_d-2m_q$). 

\begin{figure}[ht]
  \begin{center}
    \begin{tabular}{cc}
      \resizebox{65mm}{!}{\includegraphics{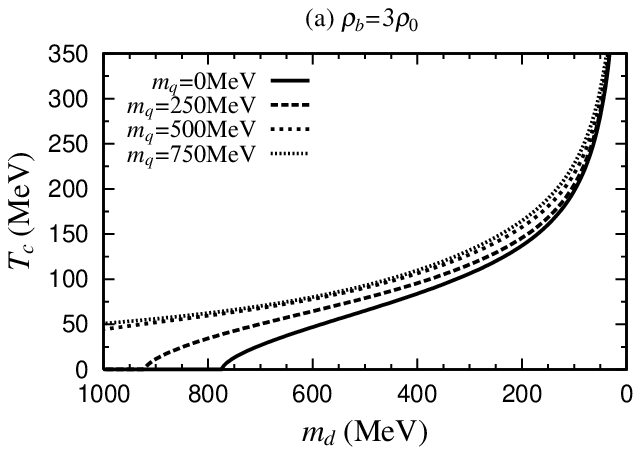}} &
      \resizebox{65mm}{!}{\includegraphics{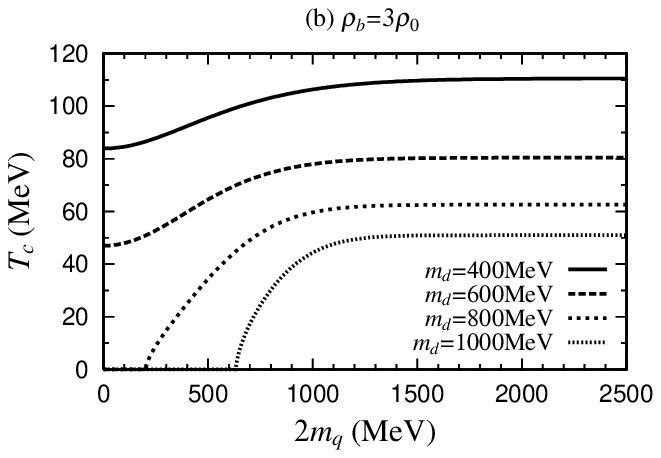}}
    \end{tabular}  
\caption{Transition temperature $T_C$ of diquark BEC:
         diquark-mass ($m_d$) dependence (a) 
         and quark-mass ($m_q$) dependence (b),
         at $\rho_b =3 \rho_0$.
         Several lines are for different quark masses 
         $m_q =0 \sim 750\,{\rm MeV}$ (a), 
         and for diquark masses $m_d =400 \sim 1000 \,{\rm MeV}$ (b).}
 \label{fig2}  
  \end{center}
\end{figure}

In the very small mass region of diquark ($m_d \sim 0$), 
which corresponds to the extremely tight binding case 
(strong-coupling limit), 
we find rapid increases of $T_C$,
which is divergent at the massless limit $m_d \rightarrow 0$
(Fig.~\ref{fig2}a). 
It can be understood 
from the range (\ref{conditionmud1}) 
and the condition (\ref{QCD_chemical_eq4}) of the chemical potentials, 
from which we obtain $\mu_q=\mu_d=0$ in the present limit, meaning that
quark and thermal diquark components of the baryon-number densities vanish
($n_q-n_{\qB}=0$, $n_d^{(th)}-n_{\dB}^{(th)}=0$) and the conserved
baryon-number density is fully compensated by the condensed diquarks
($n_b=2\cdot3\cdot n_d^{(0)}$) at any temperature 
(that is, $T_C\rightarrow \infty$).
This peculiar feature is characteristic of the massless boson 
with the conserved quantity 
which is not satisfied by photons, et al..
% 
%--------------
%----------------
%
Quark mass dependences of $T_C$ are shown in Fig.~\ref{fig2}b,
and we can find that, for large values of $m_q$, 
$T_C$ converges into a finite limit 
that depends on diquark mass $m_d$:
$T_C$ of free bose gas with mass $m_d$.
It is because the quark degrees of freedom are suppressed 
in the limit of large quark mass. 

%%%%%%%%%%%%%%%%%%%%%%%%%%%%%%%%%%%%%%%%%%%%%%%%%
\subsection{Mass phase diagram of quark-diquark system}
%%%%%%%%%%%%%%%%%%%%%%%%%%%%%%%%%%%%%%%%%%%%%%%%%

\begin{figure}[ht]
  \begin{center}
      \resizebox{70mm}{!}{\includegraphics{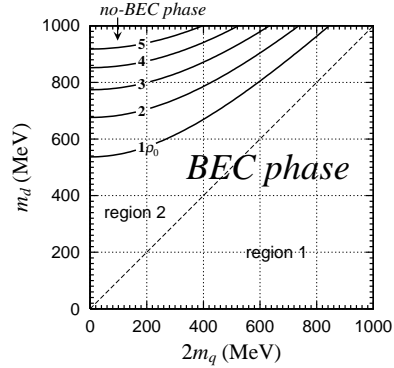}} 
     \caption{Mass phase diagram of diquark BEC for $\rho_b=(1\sim 5)\rho_0$.
       Curved lines correspond to critical borders 
       between BEC and no-BEC phases.
       Dashed line is for $2m_q=m_d$, separating region 1 (bound
   state case ; $m_d < 2 m_q$) and region 2 (resonant state case ; $m_d > 2 m_q$).}
 \label{mass_phase_diagram}
  \end{center}
\end{figure}

In the previous subsection, 
numerical results (Fig.\ref{fig1}) show 
that there exists the region in $(2 m_q,m_d)$-plane 
where no BEC occurs.
In Fig.~\ref{mass_phase_diagram}, 
we plot $T_C=0$ lines in $(2 m_q,m_d)$-plane 
for several baryon-number densities $\rho_b(=n_b/3)$. 
This line is the border of the BEC and 
no-BEC phases (below and above the line),
and we call it mass phase diagram, 
where BEC phase is defined with finite $T_C$. 
 
%
%--------------
%----------------

Now we show an analytical derivation 
of the border in the mass phase diagram 
by taking the $T_C \rightarrow 0$ limit in the formula.
At $T=0$, all antiparticles vanish, 
so that eq.~(\ref{QCD_cons1}) becomes
\begin{eqnarray}
     n_b=\ldk 3\cdot 2\cdot 2\cdot n_q
        + 2\cdot 3\cdot 1\cdot 1\cdot n_{d} \rdk_{T \rightarrow 0}.
\label{QCD_cons1_limit}
\end{eqnarray}
The diquark density generally consists 
of the condensed and thermal part 
$n_d =n_d^{(0)} +n_d^{(th)}$.
Just at $T=T_C$, 
the chemical potential reaches its upper limit 
$\mu_d =m_d$ 
but there is still no BEC component,
and thermal diquarks are completely frozen in the limit of $T=0$, 
so that we obtain $n_d =0$ at $T=T_C=0$.
The quark density is a monotonously-increasing function 
of $\mu_q$, 
and it takes the maximum value at $T =T_C =0$;
$n_q^{{\rm max}}$ with $\mu_q =\mu_d/2 =m_d/2$. 
Thus  eq.~(\ref{QCD_cons1_limit}) becomes
\begin{equation} 
     n_b =3\cdot 2\cdot 2\cdot n_q^{{\rm max}}.
\label{massphase0}
\end{equation}
From eq.~(\ref{fermi2}), 
the $n_q^{{\rm max}}$ is evaluated to be
\begin{eqnarray}
     n_q^{{\rm max}} &=&\ldk \frac{1}{2\pi^2}
                             \int_{m_q}^{\infty} 
                             \varepsilon
                             \sqrt{\varepsilon^2 -m_q^2}  
                             \cdot \theta (\mu_q-\varepsilon)
                             d\varepsilon 
                             \rdk_{\mu_q =m_d/2} 
\nonumber\\
                     &=& \frac{1}{6 \pi^2}
                         \ldk\lk \frac{m_d}{2} \rk^{2}
                            -\lk\frac{2 m_q}{2} \rk^{2} 
                         \rdk^{3/2} 
                         \times \theta(m_d -2 m_q). 
\label{massphase1}
\end{eqnarray}
In the case of $m_d < 2m_q$ (region 1 in Fig.~\ref{mass_phase_diagram}), 
the above equation gives $n_q^{{\rm max}} =0$
(no quarks), 
which leads to the contradictory result of $n_b =0$ with (\ref{massphase0}).
It means that $T_C > 0$ and all quarks are combined into diquarks with
manifest advantage of binding energy and
should be condensed to BEC state at $T=0$ (the BEC phase).

When $m_d > 2 m_q$ (region 2 in Fig.~\ref{mass_phase_diagram}), 
eqs.~(\ref{massphase0},\ref{massphase1}) gives 
\begin{equation}
     n_b = \frac{2}{\pi^2}
           \ldk \lk \frac{m_d}{2} \rk^2
               -\lk \frac{2 m_q}{2} \rk^2 \rdk^{3/2}. 
\label{massphase2} 
\end{equation}
For the fixed value of $n_b$, 
it gives the hyperbolic curve: 
the border between the BEC and no-BEC phases 
on $(2 m_q, m_d)$-plane 
as shown in Fig.~\ref{mass_phase_diagram}.
The behavior of this border is decided by the balance between kinetic energy loss of
isolated quarks (Pauli-blocking effect) and resonance energy loss of
composite diquarks.

%%%%%%%%%%%%%%%%%%%%%%%%%%%%%%%%%%%%%%%%%%%%%%%%
\subsection{Diquark condensations for constituent and current quarks:
transition temperature}
%%%%%%%%%%%%%%%%%%%%%%%%%%%%%%%%%%%%%%%%%%%%%%%%

Now we take a look at proposed values of the quark and diquark masses
in quark matter, given by Anselmino, et al.~\cite{AEFL3}, 
where two kinds of the quark/diquark masses have been presented:
\begin{equation}
\begin{array}{rll}
     &\hbox{case 1} : m_d =500 \,{\rm MeV}, &m_q =200 \sim 300 \,{\rm MeV}, 
\\
     &\hbox{case 2} : m_d =200 \,{\rm MeV}, &m_q =0   \sim 10  \,{\rm MeV}.
\end{array}
\label{cases}
\end{equation} 
The case 1 corresponds to 
dynamical chiral symmetry breaking 
which remains even
in the deconfinement phase, and thus
the large quark mass comes from the, as it were, constituent quarks. 
Whereas the case 2 is for the chiral restored quark matter 
with small current quark mass. 
For the diquark masses, 
it should be noticed that the diquark is bound ($m_d < 2 m_q$) or
resonant ($m_d > 2 m_q$) states depending on the value of $m_q$ in case
1, but is  only the resonant state in case 2.
In this paper, we take these two cases for the quark and diquark masses. 

\begin{figure}[ht]
  \begin{center}
    \begin{tabular}{cc}
      \resizebox{65mm}{!}{\includegraphics{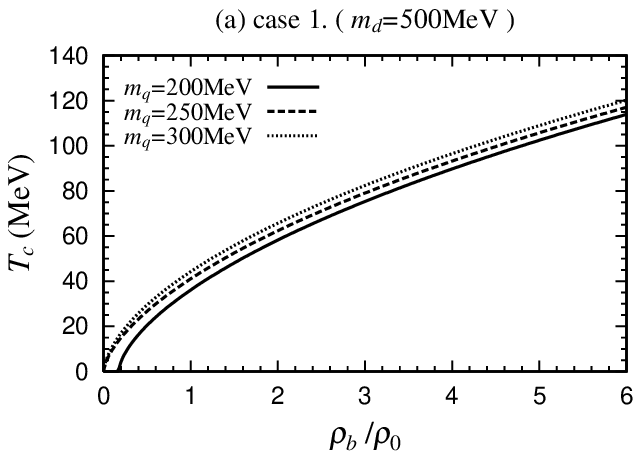}} &
      \resizebox{65mm}{!}{\includegraphics{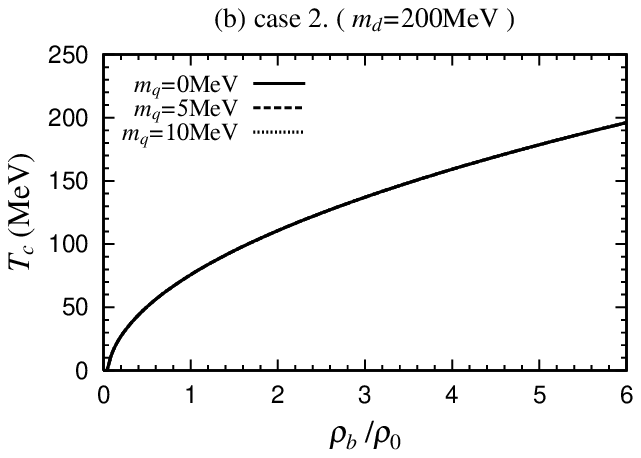}}
    \end{tabular}  
\caption{Baryon-number density ($\rho_b$) dependences of $T_C$ for case 1 (a)
         and case 2 (b).}
 \label{QCD_phase1}  
  \end{center}
\end{figure}

Fig.~\ref{QCD_phase1} shows dependence of $T_C$ on the baryon-number
density in the cases 1 and 2. 
The larger quark masses get $T_C$ higher because the
diquark binding energy ($m_d -2 m_q$) becomes larger.  
This tendency is less apparent in Fig.~\ref{QCD_phase1}b compared with
Fig.~\ref{QCD_phase1}a,
which is understood that the scale of current quark mass is too small relative to 
the energy scale of diquark BEC.

%--------------
%----------------

\begin{figure}[ht]
  \begin{center}
    \begin{tabular}{cc}
      \resizebox{65mm}{!}{\includegraphics{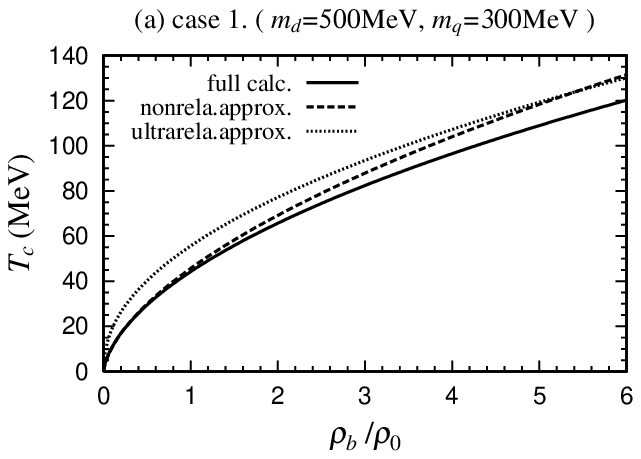}} &
      \resizebox{65mm}{!}{\includegraphics{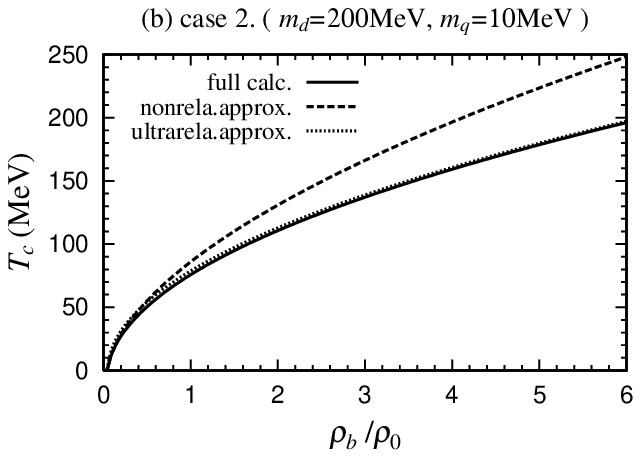}}
    \end{tabular}  
\caption{Baryon-number density ($\rho_b$) dependences of $T_C$ 
         for case 1 with $(m_d, m_q)=(500, 300)\,{\rm MeV}$ (a)
         and case 2 with $(m_d, m_q)=(200, 10)\,{\rm MeV}$  (b),
         accompanying with $T_C$ in nonrelativistic and
         ultrarelativistic limits.} 
  \label{NR_UR_1}  
  \end{center}
\end{figure}

In Fig.~\ref{NR_UR_1}, 
comparison for $T_C$ in the nonrelativistic and ultrarelativistic limits
is shown 
in the cases 1 and 2. 
%
%--------------
%----------------
%
The analytical representations of $T_C$ in these limits are given by
\begin{equation}
     T_C \sim \cases{
                3.313 \frac{n_{b,B}^{2/3}}{m_B},  & (nonrelativistic limit) \cr
                1.732 \sqrt{\frac{n_{b,B}}{m_B}}. & (ultrarelativistic limit) \cr}
\label{QCDTC}
\end{equation}     
The derivation of these formulas are given in Appendix~A.
 
In Fig.~\ref{NR_UR_1}a (case 1), 
$T_C$ is found to be
approximated by the nonrelativistic formula 
in lower density region ($\rho_b \lesssim 2 \rho_0$), 
and approaches to the ultrarelativistic one 
in the high density region; 
in the intermediate region, the gradual shift is shown between these two
limits.
It is a typical behavior of the relativistic system.
On the other hand, 
in case 2, 
the small current quark mass makes the ultrarelativistic approximation 
valid from the very small density region (Fig.~\ref{NR_UR_1}b). 

%%%%%%%%%%%%%%%%%%%%%%%%%%%%%%%%%%%%%%%%%%%%%%%%
\subsection{Diquark condensations for constituent and current quarks:
density profiles of quarks and diquarks}
%%%%%%%%%%%%%%%%%%%%%%%%%%%%%%%%%%%%%%%%%%%%%%%%

\begin{figure}[ht]
  \begin{center}
    \begin{tabular}{cc}
      \resizebox{65mm}{!}{\includegraphics{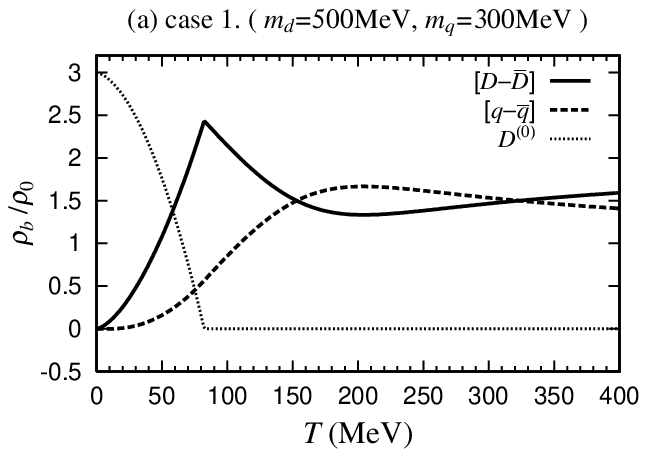}} &
      \resizebox{65mm}{!}{\includegraphics{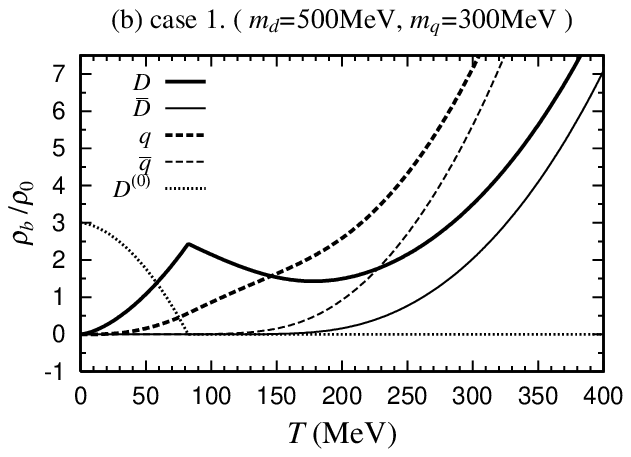}}
    \end{tabular}  
\caption{$T$-dependence of the baryon-number densities (a) 
         and the particle/antiparticle-number densities (b) 
         for constituent quarks (case 1) at $\rho_b =3 \rho_0$.
         Lines $q$ and $\qB$ are for $\rho_{q,{\bar q}}$, 
         and those $D$ and ${\bar D}$ are for $\rho_{d,{\bar d}}^{(th)}$.
         Lines $[q-\qB]$ and $[D-{\bar D}]$ are for $\rho_{b,q}$ 
         and $\rho_{b,d}^{(th)}$.
         Line $D^{(0)}$ is for the condensed part $\rho_{b,d}^{(0)} =\rho_d^{(0)}$.}
 \label{B1}  
  \end{center}
\end{figure}

\begin{figure}[ht]
  \begin{center}
    \begin{tabular}{cc}
      \resizebox{65mm}{!}{\includegraphics{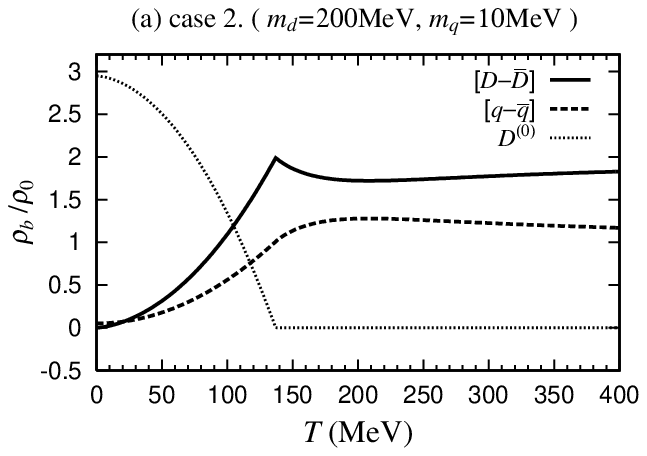}} &
      \resizebox{65mm}{!}{\includegraphics{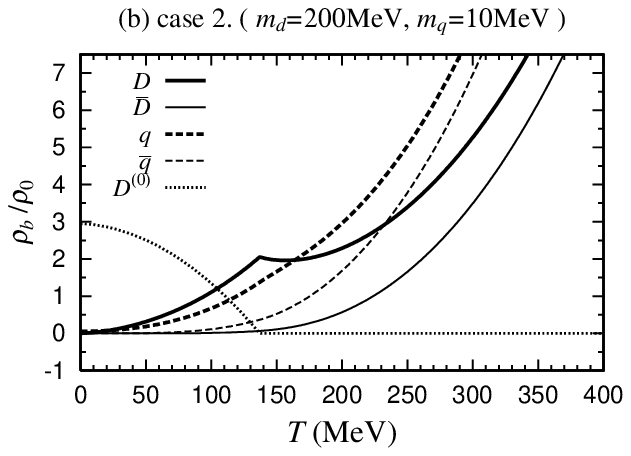}}
    \end{tabular}  
\caption{$T$-dependence of the baryon-number densities (a) 
         and the particle/antiparticle-number densities (b) 
         for current quarks (case 2) at $\rho_b =3 \rho_0$.
         For conventions, see the caption of Fig.~\ref{B1}.}
 \label{B2}  
  \end{center}
\end{figure}

Here we discuss the temperature dependences 
of particle and antiparticle densities  
shown in Fig.~\ref{B1} (case 1) 
and Fig.~\ref{B2} (case 2) at $\rho_b =3 \rho_0$.
The baryon-number density of quark 
is defined by $\rho_{b,q} =\rho_q -\rho_{{\bar q}}$, 
and that of diquark consists of the thermal part 
$\rho_{b,d}^{(th)} =\rho_d^{(th)} -\rho_{\dB}^{(th)}$ 
and the condensed part $\rho_{b,d}^{(0)} =\rho_d^{(0)}$ 
(Here we consider only the case of $\rho_b > 0$, 
so that no condensed antidiquarks exist; $\rho_{\dB}^{(0)} =0$). 

%
%--------------
%----------------
%

%
%--------------
%----------------
% 

\begin{figure}[ht]
  \begin{center}
      \resizebox{65mm}{!}{\includegraphics{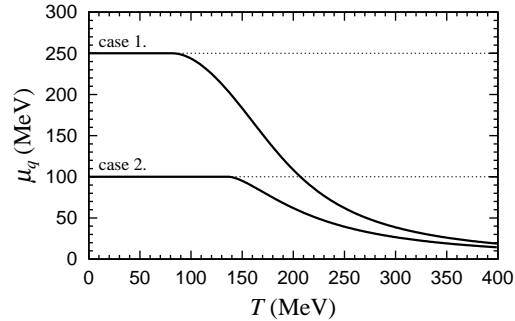}}
 \caption{Temperature dependences of quark chemical potentials
             in case 1 and 2.}  
 \label{mu_F_pro}  
  \end{center}
\end{figure}

In Figs.~\ref{B1} and \ref{B2},
$\rho_d^{(0)}$ begins to be finite at a sufficiently low temperature
showing BEC phase transition.
It should be noticed in particular that the case 2 is 
diquark BEC in the unstable resonance state 
of the current quarks (\ref{cases}),
which shows the stabilization of the resonance states in the matter
through the equilibrium processes.
The existence of the BEC is also confirmed 
in the saturation behavior of the quark chemical potential 
(Fig.~\ref{mu_F_pro});
it shows that $\mu_q (=\mu_d/2)$ reaches the upper values $m_d/2$ 
at $T_C$ 
and keeps this value in $T < T_C$. 
%
%-----------------------
%----------------

In case 1, 
the diquark is in the bound state ($m_d < 2 m_q$) 
and the association into the diquark ($q+q \rightarrow d$) 
gains the binding energy, 
which gives $\rho_{b,q} =0$ at $T\rightarrow 0$;
all quarks are combined into diquarks.
In case 2, 
the diquark formation increases the total energy of the system 
because the diquark is a resonance state ($m_d > 2 m_q$), 
and increase of the quark Fermi-energy 
(and also the entropy effect) competes with it.
Balancing these effects, 
the system does not so much favor the pairing into bosonic molecules, 
and quarks still remain even at $T\simeq 0$ (Fig,~\ref{B2}a). 
These behaviors around zero temperature is generally discussed by 
using mass phase diagram (Fig.~\ref{mass_phase_diagram}) in the previous subsection.

\begin{figure}[ht]
  \begin{center}
    \begin{tabular}{cc}
      \resizebox{65mm}{!}{\includegraphics{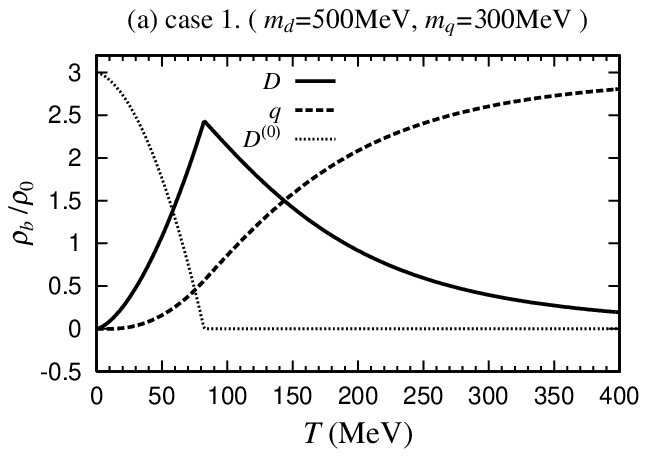}} & 
      \resizebox{65mm}{!}{\includegraphics{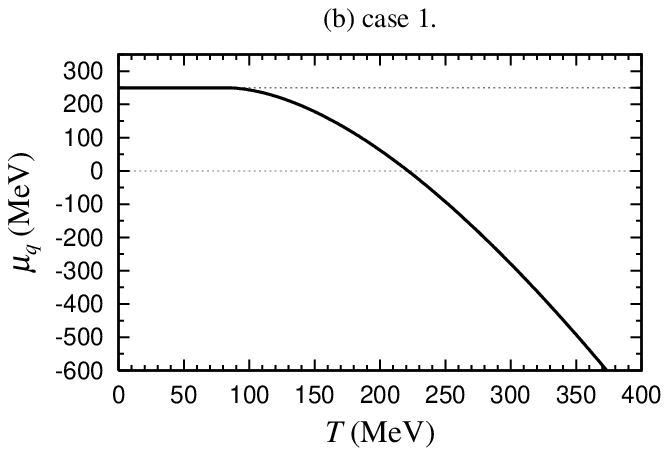}}
    \end{tabular}
 \caption{Temperature dependences of the particle-number density (a) and 
          quark chemical potential (b) 
          in case of no antiparticle degrees of freedom.
          $(m_d,m_q) =(500,300)\,{\rm MeV}$, $\rho_b=3 \rho_0$.
          For conventions, see Fig.~\ref{B1}.}  
 \label{no-anti1}  
  \end{center}
\end{figure}

Now we turn to discussion for the high-$T$ region. 
Taking case 1 first (Fig.~\ref{B1}b),
we find that the thermal diquark density $\rho_d^{(th)}$ decreases 
from $T \sim 100\,{\rm MeV} \sim (2 m_q -m_d)$, 
around which the dissociation should be effective, 
up to $T \sim 200\,{\rm MeV}$, 
but at this temperature  
the particle-antiparticle pair creations begin to occur, 
and then the diquark (and any other particle) density 
increases again.
In case 2 (Fig.~\ref{B2}b), 
we also find the density increase 
that can be attributed to the pair creation effects.
These diquark-density increases are very characteristic phenomena 
in the relativistic condensates.
Really as shown in Fig.~\ref{no-anti1}, 
in case of no antiparticle degrees of freedom,
all composite molecules should dissociate into isolated fermions
due to entropy enhancement.

%-----------------------
%----------------

Next we analyze asymptotic behaviors of the quark and diquark densities 
at the high-temperature limit. 
%
%--------------
\begin{figure}[bht]
  \begin{center}
    \begin{tabular}{cc}
      \resizebox{65mm}{!}{\includegraphics{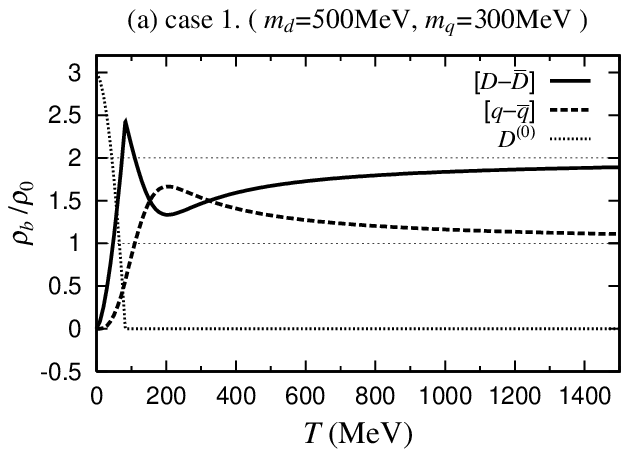}} &
      \resizebox{65mm}{!}{\includegraphics{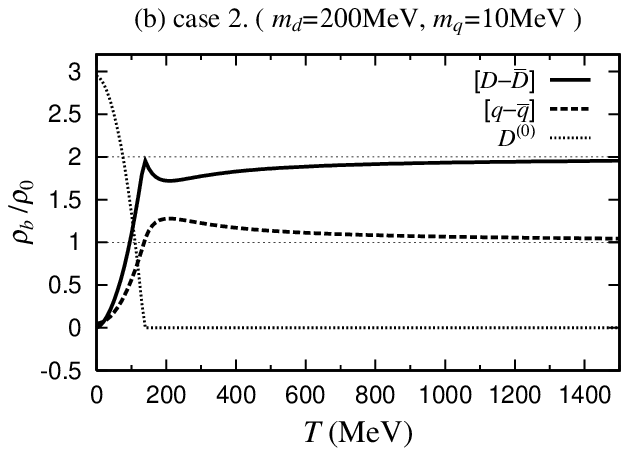}}
    \end{tabular}  
\caption{Asymptotic behaviors of the quark/diquark baryon-number densities 
         at large $T$ and $\rho_b=3\rho_0$
         for case 1 (a) and 2 (b).
         For conventions, see Fig.~\ref{B1}.}
 \label{HT_B_1}  
  \end{center}
\end{figure}
%----------------
%
In Fig.~\ref{HT_B_1}a and b, 
the high-$T$ behaviors are shown  
of the quark/diquark baryon-number densities 
in the case 1 and 2. 
From these figures, 
we can find a feature common in both cases:  
the ratio $\rho_{b,q}/\rho_{b,d} \rightarrow 1/2$
at $T \rightarrow \infty$.
From the analysis given in Appendix~B, 
we obtain a universal representation at the high-temperature limit: 
\begin{eqnarray}
     R_{q/d} \equiv \left.\frac{\rho_{b,q}}{\rho_{b,d}} \right|_{T \to \infty}
             = \lk\frac{1}{g_c}\rk^2 \lk\frac{g_q}{g_d}\rk g_{st}, 
\label{ratio2}
\end{eqnarray}
\begin{eqnarray}
    g_c\cdot\mu_q &\equiv& \mu_d,\label{ratio2_1}\\
    g_{st} &\equiv& \ltk \sum_{l=1}^{\infty}\frac{(-1)^{l-1}}{l^2} \rtk/%
                    \ltk \sum_{\acute{l}=1}^{\infty}\frac{(+1)^{\acute{l}-1}}{\acute{l}^2} \rtk%
              =1/2, \label{ratio2_2}   
\end{eqnarray}
where $g_{q,d}$ are the multiplicity factors 
for quark and diquark in the present system, 
which represent the degenerate degrees of freedom 
($\hbox{color} \times \hbox{flavor} \times \hbox{spin}$).
$g_c$ is the composite factor indicating that a diquark is constructed
by $g_c\times$quarks, 
and $g_{st}$ is the statistical factor which reflects the statistical
difference between quarks (fermions) and diquarks (bosons).
In the present case,   
$g_q=3\cdot 2\cdot 2$, $g_d=3\cdot 1\cdot 1$, $g_c=2$, and $g_{st}=1/2$.
With these values the ratio (\ref{ratio2}) becomes $R_{q/d} =1/2$,
which shows that the numerically obtained limit is universal, 
being independent of masses $m_{q,d}$.

As shown in Appendix~B, 
the quantum statistics and the antiparticle degrees of freedom are essential 
in the derivation of the correct asymptotic limit (\ref{ratio2}).
Especially 
existence of the statistical factor $g_{st}$ in (\ref{ratio2})
implies that the quantum statistics still remains 
even in the high-temperature limit.
This is understood that particles and antiparticles are 
drastically produced through the pair creation with large thermal energy, 
and Boltzmann region corresponding to high-temperature and dilute limits
almost disappears. 

Finally, we should mention consistency of the present method 
with real QCD at high temperature and density. 
Diquarks are dynamically-produced composite states 
by interquark interactions,
which should be suppressed due to the asymptotic freedom 
in the high temperature and/or density.
Furthermore,
the Pauli blocking effect on the quark-quark interaction 
prevents quark pairs from forming composite diquarks, 
and the diquarks should become correlated pairs with a coherence length
comparable with the interdiquark distance at the high density.
Such effects of increasing density for pairing mechanism is referred as `Mott effect'  
discussed with nuclear matter in refs.~\cite{SRS1, Ropk1}.
Since nuclear force has a finite repulsive core in general, 
nuclear matter should undergo superfluid-normal phase transition at a
specific density with short interparticle distance comparable to the
range of the hard core,
after the BEC-BCS crossover as density increases. 
QCD does not have such a specific scale of interaction length in the deconfinement phase, 
thus one can expect that the system is still in the superfluid phase 
even with infinite density at least around zero temperature. 
In this high density regime, however, diquarks as composite degrees
become unessential with large coherence length as discussed above. 
Thus applicability of the present method should be restricted 
in relatively low-density region, 
and incorporation of these effects in some fashion is needed for
extension to higher densities.

%%%%%%%%%%%%%%%%%%%%%%%%%%%%%%%%%%%%%%%%%%%%%%
\section{Effects of interdiquark interactions for Diquark BEC}
%%%%%%%%%%%%%%%%%%%%%%%%%%%%%%%%%%%%%%%%%%%%%%

The quarks and diquarks discussed in this paper should be 
quasiparticles in quark matter;
the interaction effects are partially included 
in the existence itself of the composite states (diquark) 
and in the effective masses 
of the quasiparticles (quark and diquark).
These effects have been included 
in the calculations in the previous subsections, 
so that they are not for the free quarks in the real meaning.
However, we have still a missing effect, 
which are also not included 
in the Gaussian-type approximations~\cite{NSR1, Rand1}, 
that the quasiparticles can scatter into the different states 
through the residual interactions.
In this subsection, 
we consider the diquark-diquark interaction  
within a mean-field approximation, 
and figure out its effects
for the transition temperature of the diquark BEC. 

As shown later, 
in the single-component boson systems, 
the boson-boson interaction effect does not change 
$T_C$ from that of the free bose gas (\ref{ApA12a}) 
within the mean-field approximation~\cite{FW1};
it is consistent with the observed $T_C$ 
of atomic BECs, 
which is rather a strongly interacting system~\cite{DGPS}.
The boson-boson correlation effects 
somewhat shift the value of $T_C$ 
in the observed $\lambda$-transition 
of the liquid $\mbox{He}^4$: $T_\lambda =2.17\,{\rm K}$ 
(eq.~(\ref{ApA12a}) gives $T_C =3.1\,{\rm K}$), 
which is also shown 
in theoretical estimations~\cite{FW1, Kapus1, HW1, BB1, BB2}.
In the diquark BEC, 
differently from the single-component boson system,
the interaction effects are found to be effective 
for $T_C$ 
still in the mean-field approximation 
through the chemical equilibrium process ($q+q \leftrightarrow d$), 
so that the calculations in this level give a first-order approximation.

We start with a model of interacting color-${\bar {\bf 3}}$ scalar fields 
$\phi_{\alpha}$ ($\alpha=1,2,3$)~\cite{Kapus1, HW1}, 
which represents the diquark degree of freedom:
\begin{eqnarray} 
     L_{\rm{eff}} =  \partial_\mu \phi^{\dagger} 
                       \partial^\mu \phi
                      -m_d^2 \phi^\dagger  \phi
                  -\lambda\lk \phi^\dagger \phi \rk^2, 
\label{Lag1}
\end{eqnarray} 
where $\phi^4$-term gives a diquark-diquark scattering effect 
(the Gross-Pitaevski approach). 
Here we have assumed
that the higher-order scattering terms ($\phi^{6,8,\dots}$) are renormalized 
into the two-body interaction~\cite{Jack1}. 
It should be noted that the Ginzburg-Landau free-energy for the CSC phase 
also has a similar form~\cite{IB1, NSY1}. 

In the mean-field approximation, 
the interaction energy of the system is given by
the expectation value of the interaction term in (\ref{Lag1}): 
\begin{eqnarray}
     \langle H_I \rangle 
          &\equiv& \langle \lambda \int d^3{\bf x}
                           \phi_{\alpha}^{\dagger}({\bf x}) 
                           \phi_{\alpha}({\bf x}) 
                           \phi_{\beta}^{\dagger}({\bf x})
                           \phi_{\beta}({\bf x}) 
                    \rangle 
\nonumber\\ 
          &\rightarrow&    \lambda \int d^3{\bf x} 
                           \langle \phi_{\alpha}^{\dagger}({\bf x}) 
                                   \phi_{\alpha}({\bf x}) \rangle 
                           \langle \phi_{\beta}^{\dagger}({\bf x})
                                   \phi_{\beta}({\bf x}) \rangle 
\nonumber\\
          &=& \frac{\lambda}{4V}
              \sum_{{\bf k},{\bf l},\alpha,\beta}
              \ldk 
              \frac{n_{{\bf k}}^\alpha n_{{\bf l}}^\beta           }{
                    \omega_{{\bf k}}^\alpha \omega_{{\bf l}}^\beta }
             +\frac{2 n_{{\bf k}}^\alpha {\bar n}_{{\bf l}}^\beta  }{
                    \omega_{{\bf k}}^\alpha \omega_{{\bf l}}^\beta }
             +\frac{{\bar n}_{{\bf k}}^\alpha {\bar n}_{{\bf l}}^\beta }{
                    \omega_{{\bf k}}^\alpha \omega_{{\bf l}}^\beta     }
             +\frac{2 n_{{\bf k}}^\alpha                           }{
                    \omega_{{\bf k}}^\alpha \omega_{{\bf l}}^\beta }
             +\frac{2 {\bar n}_{{\bf k}}^\alpha                    }{
                    \omega_{{\bf k}}^\alpha \omega_{{\bf l}}^\beta }
             +\frac{1}{\omega_{{\bf k}}^{\alpha} \omega_{{\bf l}}^{\beta}}
              \rdk, 
\label{int1}
\end{eqnarray}
where $n_{{\bf k}}^{\alpha}$ and $\bar{n}_{{\bf k}}^{\alpha}$ 
are the occupation numbers of the diquark and the antidiquark states 
with a color index $\alpha$ and a wave vector ${\bf k}$, 
and $\omega_{{\bf k}}^\alpha =\sqrt{{\bf k}^2 +m_d^2}$.

Now we consider the cases 1 and 2 
in (\ref{cases}) 
just above $T_C$. 
In both cases, 
the antiquark densities can be neglected 
because $T_C$ (a few tens MeV as in Figs.~\ref{QCD_phase1}) 
is smaller than the diquark pair-creation energy 
($2 m_d =400\,{\rm MeV} \sim 1\,{\rm GeV}$).
We also keep the leading-order term for the particle-number density
because we consider the high-density region~\cite{DS1}.
Thus the interaction energy (\ref{int1}) becomes 
\begin{eqnarray}
     \langle H_I \rangle 
          &\simeq& \frac{\lambda}{4V} 
                   \sum_{{\bf k}, {\bf l}, \alpha, \beta} 
                   \frac{n_{{\bf k}}^\alpha n_{{\bf l}}^\beta           }{
                         \omega_{{\bf k}}^\alpha \omega_{{\bf l}}^\beta } 
\nonumber\\
     &\rightarrow& \frac{\lambda}{4} 
                   \ldk V 
                   \int\frac{d^3{\bf k} d^3{\bf l} }{ (2\pi)^6 } 
                        \frac{n_{{\bf k}} n_{{\bf l}}        }{
                              \omega _{\bf k} \omega_{\bf l} } 
                  +\int\frac{d^3{\bf k}}{(2\pi)^3}
                        \frac{2 n_0 n_{{\bf k}}    }{
                              m_d \omega_{{\bf k}} }
                  +\frac{1}{V} \left( \frac{n_0}{m_d} \right)^2 
                   \rdk, 
\label{int4}
\end{eqnarray}
where $n_{{\bf k}} \equiv\sum_\alpha n_{{\bf k}}^{\alpha}$, 
and in the last line 
we have taken the continuum limit. 
It should be noted that, 
to manage the BEC singularity, 
we picked up the $\bk =0$ term $n_0$ from the ${\bf k}$ integrals
in the continuum limit in (\ref{int4}).
Eventually, 
the total energy of the system is given by
\begin{equation}
     E_d = V \int\frac{d^3{\bf k}}{(2\pi)^3} 
             \omega_{{\bf \bk}}
             ( n_{{\bf k}} +{\bar n}_{{\bf k}} )
           +\langle H_I \rangle, 
\label{total_E1}
\end{equation}
Differentiating with respect to $n_{{\bf k}}$, 
we obtain the single-particle energy:
\begin{eqnarray} 
     \varepsilon_{{\bf k}} &=& \frac{(2\pi)^3}{V}
                               \frac{\delta E_d}{\delta n(k)}
                            =  \sqrt{{\bf k}^2+m_d^2}
                              +\frac{B}{\sqrt{{\bf k}^2+m_d^2}}, 
\label{single_E1}\\
     B &\equiv& \frac{\lambda}{2} \ldk 
                \int\frac{d^3{\bf k}}{(2\pi)^3} 
                \frac{n_{{\bf k}}}{\sqrt{{\bf k}^2 +m_d^2}} 
               +\frac{n_0}{m_d V} \rdk. 
\label{BB} 
\end{eqnarray}
To evaluate (\ref{BB}), 
we concentrate on case 
of a small coupling constant  
and a large diquark mass; $B < m_d^2$. 
Generally we can say that this condition should be satisfied for small
value of $\lambda$, but its exact evaluation is not easy so that we take
it as a working hypothesis in the present paper. 
Under the condition $B < m_d^2$, 
eq.~(\ref{single_E1}) has its minimum at ${\bf k}=0$. 
More general arguments including wide range of $\lambda $ and $m_d$ will
be made elsewhere.

As discussed in the previous subsection, 
for the sufficiently larger diquark mass $m_d$,
the nonrelativistic approximation should be valid 
at the temperature close to $T_C$. 
In such a case, 
the occupation number $n_{{\bf k}}$ 
(the Bose distribution function) 
should vanish for $k \gg m_d$, 
so that we can put 
$\sqrt{{\bf k}^2 +m_d^2} \sim m_d$ 
in the first term of (\ref{BB}) 
in good approximation.
Thus, we obtain the result:
\begin{eqnarray}
     \varepsilon_{{\bf k}} &\rightarrow& 
          \sqrt{{\bf k}^2 +m_d^2} 
         +\frac{\lambda}{2m_d^2}
          \ldk \int\frac{d^3{\bf k}}{(2\pi)^3} 
                    n_{{\bf k}}
              +\frac{n_0}{V} \rdk 
\nonumber\\ 
          &=& \sqrt{{\bf k}^2+m_d^2}
             +\frac{\lambda}{2m_d^2} 
              3 \left(n_d^{(th)} +n_d^{(0)} \right),  
\label{single_E2} 
\end{eqnarray}
where $n_d^{(th)}$ and $n_d^{(0)}$ are 
the thermal and condensed densities of diquark.
It should be noted 
that the antidiquark density ${\bar n}_{{\bf k}}$ 
has been dropped in (\ref{single_E2}) 
because it is less effective 
than that of diquark 
in the nonrelativistic approximation.   

In the mean-field approximation, 
the single-particle energy appears, 
e.g., in the distribution functions 
as $\varepsilon_{{\bf k}} -\mu_d$,  
where the chemical potential $\mu_d$ is introduced 
as a Lagrange multiplier to the conserved baryon-number
$Q=\int d^3 {\bf x} j_{0}({\bf x})$
with the baryon-number current
$j_\mu=i(\phi^{\ast}\partial_\mu\phi-\phi\partial_\mu\phi^{\ast})$
for the diquark degree of freedom.
Using eq.~(\ref{single_E2}), 
we obtain
\begin{eqnarray}
     \varepsilon_{\bf k} -\mu_d 
          &=& \sqrt{{\bf k}^2 +m_d^2}
             +\frac{\lambda}{2m_d^2} 
              3 \left(n_d^{(th)} +n_d^{(0)} \right) 
             -\mu_d 
\nonumber\\ 
          &=& \sqrt{{\bf k}^2 +m_d^2} -\mu_d^*,
\label{E_chem1}
\end{eqnarray}
where the first term corresponds to the free-particle dispersion and the effective chemical potential $\mu_d^*$ 
is defined by
\begin{equation} 
     \mu_d^* \equiv \mu_d 
                   -\frac{\lambda}{2m_d^2} 
                    3 \left(n_d^{(th)} +n_d^{(0)} \right),
\label{E_che2}
\end{equation}
which is a procedure of the chemical-potential renormalization~\cite{FW1}.

With eq.~(\ref{E_chem1}), 
the boson distribution function (\ref{QCD_bosedense}) 
becomes
\begin{equation}
     n_d =m_d^3 B_R(-\beta \mu_d^*,m_d \beta).
\label{Int_bosedense}
\end{equation}

In this case, 
$T_C$ is determined by the condition $\mu_d^* =m_d$
in (\ref{Int_bosedense}): 
$n_d =m_d^3 B_R(-\beta m_d,m_d \beta)$, 
so that the formula for $T_C$ itself with fixed $n_d$ 
is completely the same as in the non-interacting bose gas;
therefore $T_C$ of single bose gas should not be changed at all through the chemical-potential renormalization~\cite{FW1}
  
Since we are considering the multi-component system  
with the chemical equilibrium ($q+q \leftrightarrow d$), 
the equilibrium condition (\ref{QCD_chemical_eq4}) 
must hold for the bare chemical potentials $\mu_{q,d}$: 
\begin{eqnarray}
     2\mu_q =\mu_d
            =\mu_d^*
            +\frac{\lambda}{2m_d^2} 
             3\left(n_d^{(th)} +n_d^{(0)} \right). 
\label{E_chemi_eq1}
\end{eqnarray}
In the equilibrium, 
this equation and the density constraint (\ref{QCD_cons1}) 
should be solved with 
the Bose and Fermi distribution,
 (\ref{Int_bosedense}) and (\ref{QCD_fermidense}).
The transition temperature $T_C$
is obtained by putting further condition $\mu_d^*=m_d$. 
It is clear that, 
differently from the single-component case, 
the interaction term in (\ref{E_chemi_eq1})
gives a $\lambda$-dependence in $T_C$,
which is the unique property of equilibrium system.

\begin{figure}[th]
  \begin{center}
    \begin{tabular}{cc}
      \resizebox{65mm}{!}{\includegraphics{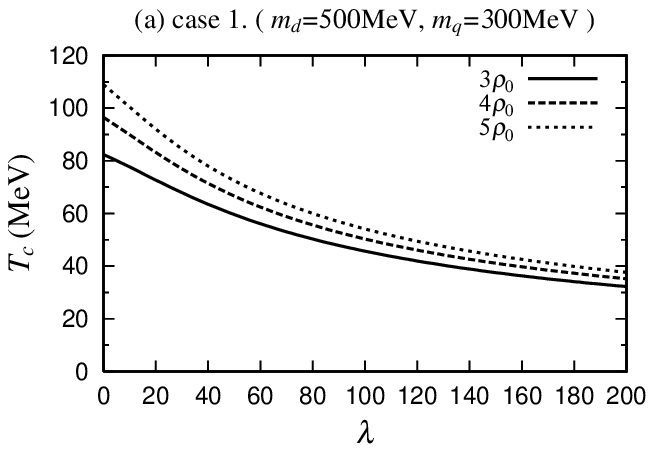}} &
      \resizebox{65mm}{!}{\includegraphics{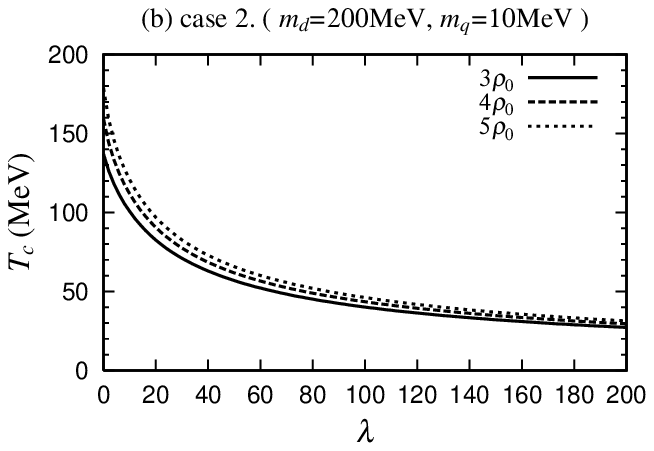}}
    \end{tabular}  
\caption{Interdiquark-interaction dependence of $T_C$
         in case 1 (a) and 2 (b) 
         for the several baryon-number densities.
         Coupling constant $\lambda$ is defined in (\ref{Lag1}).}
 \label{lambda_dep}  
  \end{center}
\end{figure}

If the interaction is repulsive ($\lambda >0$), 
it has an effect of decreasing the diquark density 
because the interaction term (\ref{single_E2}) gives 
the energy increase in the large diquark density, 
so that $T_C$ gets lower with coupling strength $\lambda$. 
This is clearly found in Fig.~\ref{lambda_dep},
where the numerical results are given  
for the $\lambda$-dependence of $T_C$ 
with the repulsive interaction
in the cases 1 and 2 in (\ref{cases}). 

%
%--------------
%----------------
%

The diquark-diquark interaction has been discussed 
for the color-${\bar {\bf 3}}$ diquarks 
by Donoghue and Sateesh~\cite{DS1}, 
where the value of $\lambda=60 \sim 100$ is estimated 
from the mass difference of the nucleon and $\Delta$  
with the assumption of the quark-hadron continuity~\cite{DS1}. 
The positiveness of $\lambda$ is also suggested in the P-matrix method~\cite{JaffLow1}.

\begin{figure}[bht]
  \begin{center}
    \begin{tabular}{cc}
      \resizebox{65mm}{!}{\includegraphics{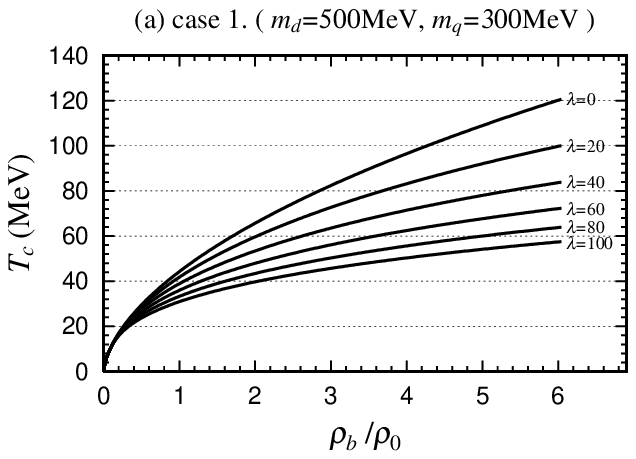}} &
      \resizebox{65mm}{!}{\includegraphics{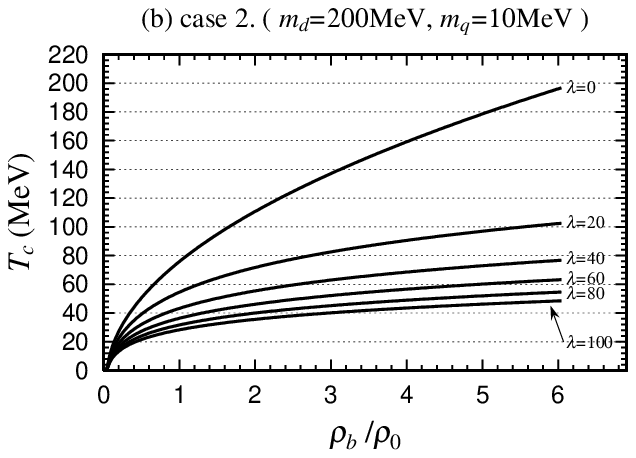}}
    \end{tabular}  
\caption{Changes of $T_C$ by the interdiquark-interaction effects 
         in case 1 (a) and 2 (b) with the baryon-number densities.
         Coupling constant $\lambda$ is defined in (\ref{Lag1}).}
 \label{phase_lambda}  
  \end{center}
\end{figure}

Figs.~\ref{phase_lambda} shows 
phase diagrams for the interacting diquark system with 
the suggested values $\lambda=60 \sim 100$.  
From these results, 
we can expect that residual diquark-diquark interactions 
have the effect to lower $T_C$
by $\sim 50\,{\rm \%}$ in comparison with that in the non-interacting cases. 
%
%--------------
%----------------

Here we should comment that 
the other interactions except the diquark-diquark one 
might give important effects, e.g.,
a possible quark-diquark interaction should generate
baryonic collective modes in relation with 
three-quark correlations.
As for these higher order correlations,
we can discuss these effects by extending our method, referring to
higher-order cluster expansion technique~\cite{SBB1}.

Finally we should discuss a possibility of the transition of quark
matter into classical liquid with large volume exclusion. Schafroth,
et al.~\cite{SBB1} pointed out that the quasi-chemical equilibrium theory
cannot be applied to the interacting fermionic system which undergoes
the condensation into
the classical liquid before the formation of difermion
molecules. To evaluate it for quark matter we now estimate the
de Boer parameter $\Lambda^{\ast}$~\cite{boer} which gives the ratio of
the quantum energy fluctuation to
the potential depth $\varepsilon$:
\begin{eqnarray}
\Lambda^{\ast}=\frac{h}{\sigma\sqrt{m\varepsilon}}=\frac{(\hbar c)
 2\pi}{\sigma\sqrt{mc^2\varepsilon}} ,\label{fluctuation1}
\end{eqnarray} 
where $\sigma$ is a range of interaction and $m$ is 
a particle mass.
Large $\Lambda^{\ast}$ corresponds to large
quantum fluctuation relative to the potential depth $\varepsilon$.
Taking typical values of several parameters for quark matter;
$\sigma\sim O(1)\mbox{fm}$, $mc^2\sim O(10)\mbox{MeV}$, $\varepsilon\sim
O(10)\mbox{MeV}$, $\Lambda^{\ast}$ is roughly estimated to be $O(10^2)$
which is very large relative to that of the quantum liquid $\mbox{He}^4$
($\Lambda^{\ast}\sim 2.64$). 
This large quantum fluctuation implies that quark matter cannot be
condensed into classical liquid 
and it must be a quantum gas or, rather, quantum liquid. 
As qualitative feature
of liquid $\mbox{He}^4$ can be well described by free bose gas model
calculations, we expect, in this sense, the quasi-chemical equilibrium
picture should be applicable for quark matter.

%%%%%%%%%%%%%%%%%%%%%%%%%%%%%%%%%%%%%%%%%%%
\section{Summary and outlook}
%%%%%%%%%%%%%%%%%%%%%%%%%%%%%%%%%%%%%%%%%%%

We have generally developed the quasi-chemical equilibrium theory 
for the system of isolated fermions and composite bosons of fermion pairs
in the relativistic framework, 
and derived the conditions for occurrence of the composite-boson BEC.
Especially a general property of BEC 
in a multi-component system is suggested;
only the lightest composite bosons can be condensed to the BEC state
(one-BEC theorem).

An application has been done for the diquark condensates 
in quark matter at moderate densities,
and only 2SC-type diquarks are treated by virtue of the one-BEC theorem. 
In this framework we have discussed the color superconductivity as the
diquark BEC in a possible BCS-BEC crossover regime.
We found the mass phase diagrams 
which determine the occurrence of BEC at a certain temperature for 
various mass values of quarks and diquarks.
With phenomenological values of model parameters, we obtain
$T_C$ as a few tens MeV, 
which is lower than those by model calculations 
within mean-field approximations, 
$\sim O(100)\,{\rm MeV}$. 
In any case, 
$T_C$ has a large dependence on quark  
and diquark masses, and it varies from $0$ to $\infty$ 
along with $m_d\rightarrow 0$ (strong coupling limit).

At the high-$T$ limit, 
antiparticles produced through the pair-creation processes 
suppress the dissociation of diquarks into quark pairs, 
and the ratio of the number densities are found to approach 
a universal value $R_{q/d} =1/2$, 
which reflects that quantum statistics still remains 
even in the high-$T$ limit.

We have also discussed effects of the diquark-diquark interaction,
and found within the mean-field approximation 
that a possible repulsive interaction lowers $T_C$ by $\sim 50\,{\rm \%}$
in comparison with that of non-interacting case 
through a chemical-potential renormalization. 

The other correlation effects beyond the mean-field approximation,
e.g., quark-diquark and three-quark correlations, 
should also be important in the BCS-BEC crossover problem. 
Combining the method of the Beth-Uhlenbeck approach~\cite{SRS1} 
with the present quasi-chemical equilibrium theory, 
we should discuss the correlation effects and the crossover problem 
from less model-dependent point of view. 
The study along this line is now ongoing 
and will be presented near future.

%%%%%%%%%%%%%%%%%%%%%%%%%%%%%%%%%%%%%%%%%%%
\section*{Acknowledgements}
%%%%%%%%%%%%%%%%%%%%%%%%%%%%%%%%%%%%%%%%%%%
The authors thank T. Kunihiro, T. Hatsuda, H. Suganuma, T. Tatsumi, and
J.~P. Blaizot for many useful discussions. K.N. is thankful to K. Tsumura
and T. Kojo for their critical suggestions and encouragements.

%%%%%%%%%%%%%
\appendix
%%%%%%%%%%%%%%%%%%%%%%%%%%%%%%%%%%%%%%%%%%%%%%%%%%%%
\section{Analysis of BEC-Transition temperature \label{AppB}} 
%%%%%%%%%%%%%%%%%%%%%%%%%%%%%%%%%%%%%%%%%%%%%%%%%%%%

In this appendix, 
we derive some formulas on the critical temperature of the BEC transition
for single bose gas 
in the three limiting cases: 
1) nonrelativistic limit, 
2) ultrarelativistic limit with a mass correction, and 
3) ultrarelativistic limit with negative-charge particles (antibosons). 

Here we consider the function $B_R$ defined in (\ref{bose3}), 
the denominator of which can be expanded as
\begin{eqnarray}
     B_R(-\betaT \muT_B,\betaT) 
          &\equiv& \frac{1}{2 \pi^2} 
                   \int_0^\infty
                   \frac{\cosh\theta (\sinh\theta)^2}{ 
                   e^{\betaT (\cosh\theta -\muT_B)} -1}
                   d\theta  \nonumber\\
          &=& \frac{1}{2 \pi^2} \sum_{l=1}^\infty e^{l\betaT\muT_B} 
                   \int_0^\infty
                   e^{-l\betaT\cosh\theta} 
                   \cosh\theta (\sinh\theta)^2
                   d\theta, \label{ApA1}
\end{eqnarray}  
where $\muT_B =\mu_B/m$.
The integral in the above expansion is evaluated to be
\begin{eqnarray}
     \int_0^\infty e^{-l\betaT\cosh\theta} 
                   \cosh\theta (\sinh\theta)^2
                   d\theta 
          &=& \frac{1}{4} \int_0^\infty e^{-l\betaT\cosh\theta} 
                               [ \cosh(3\theta) -\cosh\theta ] d\theta 
\nonumber\\
          &=& \frac{1}{4} \left\{ K_3(l\betaT) -K_1(l\betaT) \right\} 
           = \frac{1}{l\betaT} K_2(l\betaT), 
\label{ApA2}
\end{eqnarray}
where $K_\nu(z)$ is the modified Bessel function:
\begin{equation}
     K_\nu(z) =\int_0^\infty e^{-z\cosh\theta} \cosh(\nu\theta) d\theta,
\label{ApA3}               
\end{equation}
and we have used a formula: $K_{\nu-1}(z) -K_{\nu+1}(z) =-(2\nu/z) K_\nu(z)$~\cite{GR}. 
Substituting eq.~(\ref{ApA2}) into (\ref{ApA1}), 
We obtain
\begin{equation}
     B_R(-\betaT \muT_B,\betaT) =\frac{1}{2\pi^2}
                               \sum_{l=1}^\infty \frac{e^{l\betaT\muT_B}}{l\betaT}
                                              K_2(l\betaT).
\label{ApA4}
\end{equation}

The transition temperature $\betaT_C =m_B/T_C$ 
is obtained from eq.~(\ref{bose2}) with $\muT_B =\mu_B/m_B =1$, 
so that 
\begin{equation}
     n_B =m_B^3 B_R(-\betaT_C,\betaT_C) =\frac{m_B^3}{2\pi^2}
                               \sum_{l=1}^\infty \frac{e^{l\betaT_C}}{l\betaT_C}
                                              K_2(l\betaT_C).
\label{ApA5}
\end{equation}
Using the asymptotic expansion of $K_\nu(z)$ for $z \rightarrow \infty$~\cite{GR}:
\begin{equation}
     K_\nu(z) \sim \sqrt{\frac{\pi}{2z}} e^{-z} 
                   \sum_{k=0}^\infty \frac{\Gamma\left( \nu +k+\frac{1}{2} \right) }{
                                        \Gamma\left( \nu -k+\frac{1}{2} \right) }
                                  \frac{(2z)^{-k}}{k!},
\label{ApA6}
\end{equation}
we obtain the asymptotic expansion of (\ref{ApA5}) for large $\betaT_C$ (small $T_C$):
\begin{equation}
     n_B =\frac{1}{\lambda_{T_C}^{3}}
        \sum_{k=0}^\infty \frac{\Gamma\left( \frac{5}{2} +k \right) }{
                             \Gamma\left( \frac{5}{2} -k \right) }
                       \frac{\zeta\left(k+\frac{3}{2}\right)}{k!}
                       (2\betaT_C)^{-k},
\label{ApA7}
\end{equation}
where $\lambda_T =\sqrt{2\pi \betaT/m_B^2}$ is the thermal de~Broglie wave length 
and the $\zeta$-function is defined by $\zeta(z) =\sum_{l=1}^\infty l^{-z}$.
Taking the first term in the right-hand side of (\ref{ApA7}), 
we obtain the nonrelativistic limit for $T_C$~\cite{Lond1}:
\begin{equation}
     n_B \lambda_{T_C}^{3} \sim \zeta(3/2) \sim 2.612.
\label{ApA8}
\end{equation}

To obtain the ultrarelativistic limit, 
we use the asymptotic expansion of $K_n(z)$ with positive integer $n$  
for $z \to 0$:
\begin{equation}
      K_n(z) \sim \frac{1}{2} \sum_{k=0}^{n-1} (-1)^k 
             \frac{(n-k-1)!}{k!}
             \left(\frac{z}{2}\right)^{2k-n}.
\label{ApA9}
\end{equation}
Taking the first term of the above expansion 
and using it for (\ref{ApA5}), 
we obtain the density formula for small $\betaT_C$ :
\begin{equation}
     n_B \sim \frac{m_B^3}{\pi^2} 
            \sum_{l=1}^\infty \frac{e^{l\betaT_C}}{(l\betaT_C)^3} 
       = \frac{m_B^3}{\pi^2 \betaT_C^3} B_3(-\betaT_C),
\label{ApA10}
\end{equation}
where $B_\alpha(\nu)$ is the Appell function:
\begin{equation}
     B_\alpha(\nu) =\frac{1}{\Gamma(\alpha)} 
                    \int_0^\infty\frac{z^{\alpha-1} dz}{ e^{z+\nu} -1}
                   =\sum_{l=1}^\infty \frac{e^{-l\nu}}{l^{\alpha}}.
\label{ApA11}
\end{equation}
Using the expansion $B_3(-\betaT_C) \sim \zeta(3) +\zeta(2) \betaT_C$ 
for $\betaT_C \to 0$ (high-$T_C$ limit) in eq. (\ref{ApA10}),
we obtain the ultrarelativistic limit for $T_C$:
\begin{eqnarray}
     n_B &\sim& \frac{m_B^3}{\pi^2} \left[ \frac{\zeta(3)}{\betaT_C^3} 
                          +\frac{\zeta(2)}{\betaT_C^2} \right] 
\nonumber\\
       &=& \frac{\zeta(3)}{\pi^2} T_C^3 
        +m_B \frac{\zeta(2)}{\pi^2} T_C^2,
\label{ApA12}
\end{eqnarray}
where $\zeta(2) =\pi^2/6 \sim 1.6449$ and $\zeta(3) \sim 1.2020$.
It should be noted that the first term in the above equation does not include $m_B$;
in the ultrarelativistic limit, the boson mass is neglected 
and the energy-momentum dispersion relation is approximated to be 
$\varepsilon \sim |{\bf k}|$.
The second term in eq. (\ref{ApA12}) corresponds to the small mass correction.

In summary, we obtain the formulas of $T_C$ of the relativistic BEC:
\begin{equation}
      T_C \sim \cases{
                 3.313 \frac{n_B^{2/3}}{m_B}, & (nonrelativistic limit) \cr
                 2.017 n_B^{1/3}.             & (ultrarelativistic limit) \cr}
\label{ApA12a}
\end{equation}

Finally we consider the boson-antiboson system, 
where the similar calculation with (\ref{ttba}) gives the expansion for baryon-number density :
\begin{equation}
     n_{b,B} =n_B -n_{{\bar B}} 
             =\frac{m_B^3}{\pi^2} 
              \sum_{l=1}^\infty \frac{\sinh(l\betaT \muT_B)}{l\betaT}
                             K_2(l\betaT). 
\label{ApA13}
\end{equation}
The formula for $T_C$ is obtained by putting $\muT_B =1$ and $\betaT =\betaT_C$.
Using eq.~(\ref{ApA6}), we obtain the asymptotic expansion for large $\betaT_C$:
\begin{equation}
     n_{b,B} =\frac{1}{\lambda_{T_C}^{3}}
        \sum_{k=0}^\infty \frac{\Gamma\left( \frac{5}{2} +k \right) }{
                             \Gamma\left( \frac{5}{2} -k \right) }
                       \frac{1}{k!}
                       \left\{ \zeta\left(k+\frac{3}{2}\right) 
                              -B_{k+3/2}(2\betaT_C) \right\}
                       (2\betaT_C)^{-k},
\label{ApA14}
\end{equation}
where $B_{k+3/2}$, defined by (\ref{ApA11}), corresponds to the antiboson contributions.
In the nonrelativistic limit (low $T_C$), 
$B_{k+3/2}(2\beta_C)$ can be neglected  
because $B_\alpha(z) \sim e^{-z}$, 
so that we obtain the same formula as eqs.~(\ref{ApA7},\ref{ApA8})
where $n_B$ is replaced with $n_{b,B}$.
This means that the antiboson degree of freedom is suppressed 
and we can put $n_{b,B} \sim n_B$ in this limit.

In the ultrarelativistic limit (small $\betaT_C$), 
using (\ref{ApA9}), 
eq.~(\ref{ApA13}) becomes
\begin{equation}
     n_{b,B} \sim \frac{2 m_B^3}{\pi^2} \zeta(2) \frac{\muT_B}{\betaT^2}
             =\frac{2}{\pi^2} \zeta(2) \mu_B T^2,
\label{ApA15}
\end{equation}
where $\sinh{z}/z \sim 1$ has been used.
Substituting $\muT_B=1$ and $\betaT=\betaT_C$, 
we obtain the ultrarelativistic formula for $T_C$:
\begin{equation}
     n_{b,B} \sim \frac{2 m_B^3}{\pi^2} \frac{\zeta(2)}{\betaT_C^2}
             = 2 m_B \frac{\zeta(2)}{\pi^2} T_C^2.
\label{ApA16}
\end{equation}
Comparison with eq.~(\ref{ApA12}) shows that the leading terms from $n_{B,{\bar B}}$ 
cancels out so that the result depend on $m_B$ in spite of the ultrarelativistic limit.

In summary, we obtain the formulas of $T_C$ for the boson-antiboson system:
\begin{equation}
     T_C \sim \cases{
                3.313 \frac{n_{b,B}^{2/3}}{m_B},  & (nonrelativistic limit) \cr
                1.732 \sqrt{\frac{n_{b,B}}{m_B}}. & (ultrarelativistic limit) \cr}
\label{ApA16a}
\end{equation}     

%%%%%%%%%%%%%%%%%%%%%%%%%%%%%%%%%%%%%%%%%%%%%%%%%%%%
\section{Asymptotic form of density profiles \label{AppC}}
%%%%%%%%%%%%%%%%%%%%%%%%%%%%%%%%%%%%%%%%%%%%%%%%%%%%

First we calculate the asymptotic expansion of fermion-number density;
calculations go in the same way as for the boson density formula (\ref{ApA13}):
\begin{equation}
     n_{b,F} =n_F -n_{{\bar F}} 
             =\frac{m_F^3}{\pi^2} 
              \sum_{l=1}^\infty (-1)^{l-1} \frac{\sinh(l\betaT \muT_F)}{l\betaT}
                             K_2(l\betaT), 
\label{ApB1}
\end{equation}
where the extra factor $(-1)^{l-1}$ comes from the factor $+1$ 
in the denominator of the fermion-density (\ref{fermi3}) reflecting
Fermi statistics (Bose statistics gives the extra factor $(+1)^{l-1}=+1$
which is embedded in eq. (\ref{ApA13})).
Expanding the function $K_2$ and taking the leading-order term in (\ref{ApA9}),
we obtain the fermion-density formula for small $\betaT$ :
\begin{equation}
     n_{b,F} \sim \frac{2 m_F^3}{\pi^2} \frac{\zeta(2)}{2} \frac{\muT_F}{\betaT^2}
             =\frac{2}{\pi^2} \frac{\zeta(2)}{2} \mu_F T^2,
\label{ApB2}
\end{equation}
where $\sum_{l=1}^\infty (-1)^{l-1}/l^\alpha =(1-2^{1-\alpha})
\zeta(\alpha)$ has been used
to give the factor $\zeta(2)/2$ in (\ref{ApB2}).
In case of Bose statistics, factor 
$\sum_{l=1}^\infty (+1)^{l-1}/l^\alpha =\zeta(\alpha)$
from the analysis of eq.~(\ref{ApA13}) gives $\zeta(2)$ in
(\ref{ApA15}). 
The difference between the factors $\zeta(2)$ and $\zeta(2)/2$
in eqs.~(\ref{ApA15},\ref{ApB2}) tells that 
the effect of quantum statistics still remains even in high-$T$ limit,
which is discussed in eq.~(\ref{ratio2}).

Now we apply the boson and fermion density-formulas for high-$T$ region, 
(\ref{ApA15}) and (\ref{ApB2}), 
to quark-diquark system, 
and evaluate the high-$T$ limit of the baryon-number density ratio.
Using eqs.~(\ref{ApA15},\ref{ApB2}), 
the baryon-density ratio of isolated quarks to diquarks is given by
\begin{eqnarray}
     R_{q/d} \equiv \left.\frac{\rho_{b,q}}{\rho_{b,d}} \right|_{T \to \infty}
             =\left.\frac{(1/3) g_q n_{b,q}}{(2/3) g_d  n_{b,d}} \right|_{T \to \infty}
             = \lk\frac{1}{g_c}\rk^2 \lk\frac{g_q}{g_d}\rk g_{st}, 
\label{APratio3}
\end{eqnarray}
\begin{eqnarray}
    g_c\cdot\mu_q &\equiv& \mu_d,\label{APratio3_1}\\
    g_{st} &\equiv& \ltk \sum_{l=1}^{\infty}\frac{(-1)^{l-1}}{l^2} \rtk/%
                    \ltk \sum_{\acute{l}=1}^{\infty}\frac{(+1)^{\acute{l}-1}}{\acute{l}^2} \rtk%
              =1/2, \label{APratio3_2}   
\end{eqnarray}
where $1/3,2/3$ are the baryon numbers of quark and diquark, 
and $g_{q,d}$ are the multiplicity factors for quark and diquark in the
present system, 
which represent the degenerate degrees freedom ($\hbox{color} \times \hbox{flavor} \times \hbox{spin}$).
$g_c$ is the composite factor indicating that a diquark is constructed
by $g_c\times$quarks (in the present case, $g_c=2$), 
and $g_{st}$ is the statistical factor which reflects the statistical
difference between quarks (fermions) and diquarks (bosons).

%%%%%%%%%%%%%%%%%%%%%%%%%%%%%%%%%%%%%%%%%%%%%%%%%%%%%%%%%%%%%%%%%%%%%%%%%%%%

\end{document}